\documentstyle[amstex,12pt]{article}                 
\newlength{\dinwidth}
\newlength{\dinmargin}
\renewcommand{\theequation}{\thesection.\arabic{equation}}
\newtheorem{Definition}{Definition}[section]

\newtheorem{Proposition}[Definition]{Proposition}
\newtheorem{Lemma}[Definition]{Lemma}
\newtheorem{Corollary}[Definition]{Corollary}
\newcommand{\nin}{\noindent}

\newcommand{\CC}{{\mbox{{\small $ \Bbb C$}}}}
\newcommand{\NN}{{\mbox{{\small $ \Bbb N$}}}}
\newcommand{\RR}{{\mbox{{\small $ \Bbb R \,$}}}}
\newcommand{\II}{{\mbox{{\small $ \Bbb I$}}}}
\newcommand{\ckl}{\,{\mbox{{\scriptsize $\circ$}}}\,}
\newcommand{\RRf}{ {\mbox{{\footnotesize $ \Bbb R $}}}}
\newcommand{\KK}{ {\mbox{{\small $ \Bbb K \,$}} }}
\newcommand{\KKf}{ {\mbox{{\footnotesize $ \Bbb K$ }} }}
\newcommand{\vth}{\vartheta}
\newcommand{\la}{{\lambda}}                     
\newcommand{\Oo}{{\cal O}}                      
\newcommand{\Op}{{\cal O}'}                     
\newcommand{\Oi}[1]{{\cal O}_{#1}}              
\newcommand{\AO}{{\frak A} ({\cal O})}           
\newcommand{\AOi}[1]{{\frak A} ({\cal O}_{#1})}
\newcommand{\Al}{{\frak A }}                     
\newcommand{\fT}{\widetilde{f}}                 
\newcommand{\OT}{\widetilde{{\cal O}}}          
\newcommand{\AOT}{\widetilde{{\frak A}} ( \widetilde{{\cal O}} )}
\newcommand{\AaT}{\widetilde{A}}
\newcommand{\AT}{\widetilde{{\frak A}}}          
\newcommand{\Rl}{{R}_{\lambda}}                 
\newcommand{\Nl}{N_{\lambda}}                   
\newcommand{\AlO}{{\frak A}_{\lambda}({\cal O})} 
\newcommand{\Alu}{\underline{{\frak A}}}         

\newcommand{\Au}{\underline{A}}                  
\newcommand{\Aul}{{\underline{A}}_{\lambda}}     
\newcommand{\Pg}{{\cal P}_{+}^{\uparrow}}        
\newcommand{\Lx}{( \Lambda , x )}                
\newcommand{\ax}{{\alpha}_{x}}                   
\newcommand{\ay}{{\alpha}_{y}}                   
\newcommand{\az}{{\alpha}_{z}}                   
\newcommand{\af}{{\alpha}_{f}}                   
\newcommand{\ag}{{\alpha}_{g}}                   
\newcommand{\aL}{{\alpha}_{\Lambda}}             
\newcommand{\aLx}{{\alpha}_{\Lambda , x}}        
\newcommand{\ULx}{U(\Lambda , x)}                
\newcommand{\axu}{{\underline{\alpha}}_x}        

\newcommand{\aP}{{\alpha}_{{\cal P}_{+}^{\uparrow}}}
\newcommand{\Ho}{{\cal H}_{\omega}}              
\newcommand{\Hh}{{\cal H}}                       
\newcommand{\BH}{{\cal B}({\cal H})}             
\newcommand{\Ooo}{{{\Omega}_{\omega}}}           
\newcommand{\po}{{\pi}_{\omega}}                 
\newcommand{\oziu}{{\underline{\omega}}_{{0}, \iota}}
\newcommand{\ou}{{\underline{\omega}}}           
\newcommand{\pu}{{\underline{\pi}}}              
\newcommand{\Hziu}{\underline{\cal H}_{0,\iota}}   
\newcommand{\Omiu}{\underline{\Omega}_{0,\iota}}   
\newcommand{\pziu}{\underline{\pi}_{0,\iota}}      

\newcommand{\aLxu}{\underline{\alpha}_{{\Lambda},x}} 
\newcommand{\Dziu}{\underline{\Delta}_{0,\iota}}  
\newcommand{\Jziu}{\underline{J}_{0,\iota}}       
\newcommand{\Uziu}{\underline{U}_{0,\iota}}       
\newcommand{\Bu}{\underline{B}}
\newcommand{\Cu}{\underline{C}}
\newcommand{\Ak}{A_{\kappa}}
\newcommand{\Bk}{B_{\kappa}}
\newcommand{\auP}{\underline{\alpha}_{{\cal P}_{+}^{\uparrow}}}

\newcommand{\Hu}{\underline{\cal H}}
\newcommand{\Ou}{\underline{\Omega}}
\newcommand{\pul}{\underline{\pi}_{\lambda}}
\newcommand{\lk}{\lambda_{\kappa}}
\newcommand{\Aoi}{{\frak A}_{0,\iota}}
\newcommand{\aoi}{\alpha^{(0,\iota)}}
\newcommand{\sgl}{\sigma_{\lambda}}
\newcommand{\smu}{\underline{\sigma}_{\mu}}
\newcommand{\sglu}{\underline{\sigma}_{\lambda}}
\newcommand{\olu}{\underline{\omega}_{\lambda}}
\newcommand{\ooi}{\omega_{0,\iota}}
\newcommand{\lkk}{\lambda_{k}}
\newcommand{\oim}{0,\iota(\mu)}
\newcommand{\doim}{\delta_{\mu}^{(0,\iota)}}
\newcommand{\toim}{\tau_{\mu}^{(0,\iota)}}
\newcommand{\Rr}{{\cal R}}
\newcommand{\Wp}{{\cal W}_{+}}
\newcommand{\Wm}{{\cal W}_{-}}
\newcommand{\Uoi}{U_{0,\iota}}
\newcommand{\Omoi}{\Omega_{0,\iota}}
\newcommand{\xu}{\underline{\chi}}
\newcommand{\Iu}{\underline{I}}
\begin{document}
%
\title{{\bf \mbox{ Scaling Algebras and Renormalization} \\
\mbox{\hspace*{-25pt} Group in Algebraic Quantum Field Theory}}}
\author{Detlev Buchholz \hspace{2pt} and \hspace{2pt} Rainer Verch
 \\[6pt]
II. Institut f\"ur Theoretische Physik,
Universit\"at Hamburg \\
D-22761 Hamburg, Federal Republic of Germany}
\date{DESY 95-004 \\ hep-th/9501063   }

\maketitle
\begin{abstract}
For any given algebra of local observables in Minkowski space an
associated scaling algebra is constructed on which renormalization
group (scaling) transformations act in a canonical manner. The method
can be carried over to arbitrary spacetime manifolds and provides a
framework for the systematic analysis of the short distance
properties of local quantum field theories. It is shown that every
theory has a (possibly non-unique) scaling limit which can be
classified according to its classical or quantum nature. Dilation
invariant theories are stable under the action of the renormalization
group. Within this framework the problem of wedge
(Bisognano-Wichmann) duality in the scaling limit is discussed
and some of its physical implications are outlined.
\end{abstract}
\section{Introduction}
\setcounter{equation}{0}
The algebraic approach to relativistic quantum field theory has proven
to be an efficient setting for the structural analysis of
properties of physical systems
which manifest themselves at the upper end of the spatio-temporal
scale. Examples are the classification of the possible statistics
and superselection structure of particles, collision theory and
the clarification of the infrared properties of theories with
long range forces \cite{Ha}.
At the lower end of the scale the algebraic point of view has been less
successful, however. As
a matter of fact, basic ideas which have emerged from physics at small
scales such as the parton picture or the notion of asymptotic freedom
have not yet found an appropriate expression in the algebraic setting.
What is missing so far in this approach is the analogue of the
renormalization group, cf. \cite{Am} and references quoted there,
which allows one to transform a theory
at given scale into the corresponding theories at other
scales.

In order to understand the origin of this difficulty one has to
call to mind the conceptual foundations of the algebraic approach.
Algebraic quantum field theory is based on the idea that the
correspondence
\begin{equation}  \Oo \rightarrow \AO \end{equation}
between spacetime regions $\Oo \subset \RR^4$ and local algebras
of observables $\AO$ constitute
the intrinsic description of a theory \cite{Ha}. One therefore refrains
from assigning to the individual elements of these algebras any
particular physical interpretation. In fact, one presumes that this
interpretation is essentially fixed once the map  is given. We
adhere in the following to the standard terminology according to which
this map is called a local net.

Quantum fields, which are a basic ingredient in the conventional
approach to the renormalization group, are regarded as a kind of
coordinatization of the local algebras and therefore do not appear
explicitly in the algebraic setting. This view is justified by
the observation that different irreducible sets of field operators
which are relatively local to each other yield the same scattering
matrix \cite{Bo1}. Thus the physical content of a theory does not
depend on a particular choice of fields. All what matters is
the information about the localization properties of the operators,
i.e., the net.

The absence of quantum fields in the algebraic setting causes problems,
however, if one wants to apply the ideas of the renormalization group.
In the conventional framework of quantum field theory the
renormalization
group transformations $\Rl , \la > 0$, act on the underlying quantum
fields $\phi ( x )$ by scaling the spacetime coordinates $x$,
accompanied by a multiplicative renormalization,
$\Rl: \, \phi (x) \rightarrow \phi_\la ( x ) \doteq
\Nl \, \phi (\la x)$. One thereby maps
the theory at the original scale $\la = 1$, say, onto the corresponding
theory at scale $\la$ whithout changing the value of the fundamental
physical constants, i.e., the velocity of light $c$ and
Planck's constant $\hbar$. Moreover, by the multiplicative
renormalization factor $\Nl$, the scale of field strength is adjusted
in such a way that the mean values and mean square fluctuations of
the fields in some fixed reference state are of the same order of
magnitude at small scales. Thus the quantum fields are employed to
identify at each scale $\la$ a set of operators with a fixed physical
interpretation. These operators can then be used to compare the
properties of the theory at different scales.

It is apparent that this approach is at variance with the basic
philosophy of algebraic quantum field theory and one is faced with the
question of how to implement the renormalization
group in this setting. It is the aim of the present article to
provide a solution of
this conceptual problem and to establish a mathematical framework apt
for the structural analysis of local nets at small scales.
Our approach is based on the following elementary observations.

(i) According to the geometrical significance of the renormalization
group the transformations $\Rl$  should map the given net
$\Oo \rightarrow \AO$ at the original spatio-temporal scale $1$
onto the corresponding net $\Oo \rightarrow \AlO \doteq {\frak A}
( \la \Oo )$ at scale $\la$, i.e.,
\begin{equation} \Rl : \, \AO \rightarrow \AlO \end{equation}
for every region $\Oo \subset \RR^4$. Since space and time are scaled
in the same way the value of the velocity of light $c$ is kept fixed
under these maps.

(ii) The condition that $\hbar$ remains constant under renormalization
group transformations can be expressed in the algebraic setting as
follows. If one scales space and time by $\la$ and does not want to
change the unit of action one has to
rescale energy and momentum by $\la^{-1}$. The energy-momentum scale
can be set by determining the energy and momentum which is transferred
by the action of observables to physical states. Hence if
{}\footnote{All quantities relating to four-momentum space
$\RR^4$ will be marked by a tilde $\widetilde{\; \;}$, cf.\ the
subsequent section for precise definitions.}
$\AOT$ denotes the subspace of all (quasi local) observables which, at
the original scale $1$, can transfer energy-momentum contained in the
set $\OT \subset \RR^4$
and if $\AT_\la ( \OT ) \doteq \AT ( \la^{-1}
\OT )$ denotes the corresponding space at scale $\la$,
then the transformations $\Rl$ should induce a map
\begin{equation} \Rl : \, \AOT \rightarrow \AT_\la ( \OT )
\end{equation}
for every $\OT$. (An analogous relation should hold for the angular
momentum transfer.)

(iii) In the case of dilation invariant theories the transformations
$\Rl$ are expected to be isomorphisms, yet this will not be true in
general since the algebraic relations between observables may depend on
the scale. But since the transformations $\Rl$ are designed to
identify observables at different scales they still ought to be
continuous, bounded maps, uniformly in $\la $.
This condition is akin to the multiplicative
renormalization of fields in the conventional approach to the
renormalization group.

The above conditions subsume the physical constraints imposed on the
renormalization group transformations $\Rl$, although they do not fix
these maps. As a matter of fact, there  exists an abundance of
such maps for any given $\la > 0$. But all of these maps identify
                                       the same
net at scale $\la$,
they merely reshuffle the operators within the local algebras in
different ways. Bearing in mind the basic hypothesis of algebraic
quantum field theory according to which the physical information of a
theory is contained in the net, it should thus
not matter which map one picks for the short distance analysis of
a theory. One may consider any one of them or, what amounts to the same
thing, one may consider them all.

We adopt here the latter point of view which can conveniently be
expressed by introducing the concept of scaling algebra. Roughly
speaking, the scaling algebra consists of operator valued functions
$\la \rightarrow \Rl (A), \, \la >0$,
which are the orbits of the local observables $A$ under the
action of all admissible transformations $\Rl$. As we shall see, the
specific properties of the transformations $\Rl$ indicated above
imply that the scaling algebra still has the structure of a local
net on which the Poincar\'e group acts in a continuous manner.
Moreover, the renormalization group induces an additional symmetry of
this net: scaling transformations. Thus the physically
significant features of the renormalization group give rise to specific
algebraic properties of the scaling algebra.

The states of physical interest can be lifted to the scaling algebra
and their behaviour under scaling transformations
can then be analyzed. We will show that the transformed states have,
at arbitrarily small scales, limits which are vacuum states.
This result allows it on one hand to classify the
possible scaling limits of local nets. Besides the
cases corresponding to theories with an ultraviolet fixed point there
appear in the present general setting other possibilities, such as
theories with a classical or non-unique scaling limit. If the underlying
theory is invariant under dilations and satisfies a compactness
criterion proposed by Haag and Swieca \cite{HaSw}, it is
invariant under the action of the renormalization group and coincides
with its scaling limit.

The information that the scaling limits of physical states are always
vacuum states provides, on the other hand, the basis for a more
detailed short distance analysis. A relevant technical result in this
respect is the observation that theories satisfying a condition of
wedge duality, established by Bisognano and Wichmann \cite{BiWi}, comply
with this condition also in the scaling limit.
In an application of this result we will generalize a theorem by
Fredenhagen \cite{Fr} and show that in theories
with a non-classical scaling limit the local von Neumann algebras
$\AO^-$ corresponding to double cones $\Oo$ are of type $\mbox{III}_1$
according to the classification of Connes. This fact has
interesting physical implications, cf. for example \cite{SuWe,BuDaFr}.

The condition of wedge duality is also of vital importance in the
general
analysis of the superselection structure, carried out by Doplicher,
Haag and Roberts \cite{Ha,DHR}. Applying these methods in the present
setting one can determine
the gauge group and the particle structure appearing in the scaling
limit and thereby lay the ground for a rigorous discussion of the
parton aspects of the theory. We will deal with this issue in a
future publication.

The present article is organized as follows. In the subsequent Sec.\ 2
our assumptions are stated and some preliminary results on the momentum
space properties of local observables are derived. The scaling algebra
and the accompanying scaling transformations are
introduced in Sec.\ 3, where it is also shown how to lift physical
states to this algebra and to recover from the lifts the
physical information. This method can be generalized to nets
on arbitrary spacetime manifolds, as is outlined in the Appendix.
Section 4 contains the analysis of the scaling limits of physical
states and the resulting classification of theories. The special case
of dilation invariant theories is discussed
in Sec.\ 5, and Sec.\ 6 contains the derivation of wedge
duality in the scaling limit, as
well as a discussion of its consequences. The article concludes
with an outlook on further developments of the theory.
\section{The structure of local observables}
\setcounter{equation}{0}
For the convenience of the reader who is not familiar with the
framework of algebraic quantum field theory \cite{Ha} we briefly
list in the first part of this section the basic
assumptions and add a few comments. In the second part we establish
some  properties of local operators in momentum space which
are of relevance in the subsequent discussion.

1.  {\em (Locality)\/} We suppose
that the local observables of the underlying theory generate a local
net over Minkowski space $\RR^4$, that is an
inclusion preserving map $ \Oo \rightarrow \AO $
from the set of open, bounded regions $\Oo \subset \RR^4$ to
unital C$^*$-algebras $\AO$ on the pertinent Hilbert
space $\Hh$. Thus each $\AO$ is a norm closed subalgebra of the
algebra $\BH$ of all bounded operators on $\Hh$ which is stable
under taking adjoints and contains the unit operator, and there holds
\begin{equation} \Al (\Oi{1}) \subset \Al (\Oi{2}) \quad \mbox{if}
    \quad \Oi{1}
\subset \Oi{2}. \end{equation}
The net is supposed to comply with the principle of locality (Einstein
causality) according to which observables in spacelike separated
regions commute. In formula form,
\begin{equation} \AOi{1} \subset \AOi{2}' \quad \mbox{if} \quad  \Oi{1}
    \subset
\Oi{2}',\end{equation}
where
$\Op$
denotes the spacelike complement of
$\Oo$
and
$\AO '$
the set of operators in
$\BH$
which commute with all operators in
$\AO$.
The (global) algebra generated by all local algebras
$\AO$
(as a norm inductive limit) will be denoted by
$\Al$ and is assumed to act irreducibly on $\Hh$.

2. {\em (Covariance)\/} On $\Hh$ there exists a continuous unitary
representation $U$ of the Poincar\'e group $\Pg$ which induces
automorphisms of the net. Thus for each $\Lx \in \Pg$ there is
an $\aLx \in \mbox{Aut}(\Al)$ given by
\begin{equation} \aLx ( A ) \doteq \ULx A \ULx^{-1}, \quad A \in \Al,
\end{equation}
and, in an obvious notation,
\begin{equation} \aLx ( \AO ) = \Al ( \Lambda \Oo + x ) \end{equation}
for any region
$\Oo$.
We amend this fundamental postulate by the condition that
the operator valued functions
\begin{equation} \Lx \rightarrow \aLx (A), \; A \in \Al \end{equation}
are continuous in the norm topology. This assumption, which plays a
crucial role in the present investigation, does not impose any
essential restrictions of generality. For these functions are
always continuous in the strong operator topology, because of the
continuity properties of the representation $U$ and the fact that the
operators $A$ are bounded. Hence, by convolution
of these operator valued functions with suitable test functions, one
can always proceed to a local net
which complies with our continuity condition and which is dense in the
original net in the strong operator topology. So it still contains
the relevant information about the states of interest here.

For the sake of uniqueness we also assume that the local algebras
$\AO$ are maximal in the following sense: any operator
in the strong operator topology closure $\AO^-$ of $\AO$ which complies
with
our continuity condition is already contained in $\AO$. This condition
on the net can always be satisfied by enlarging the local algebras, if
necessary.

3. {\em (Spectrum condition)\/} The joint spectrum of the generators
of the unitary representation of the translations $U \,|\, \RR^4$ is
contained in the closed forward lightcone $\overline{V}_+$.
Moreover, there is an (up to a phase unique) unit
vector $\Omega \in \Hh$, representing the vacuum, which is invariant
under the action of the representation $U$,
\begin{equation} \ULx \Omega = \Omega, \quad \Lx \in \Pg. \end{equation}

Besides the vacuum $\Omega$ and
its local excitations, described by the vectors in $\Hh$, there exist
other states of physical interest such as charged states, thermal
states, etc. These states are not represented by vectors in $\Hh$,
but can be described by suitable positive, linear and normalized
functionals
$\omega$ on the given algebra $\Al$. By the GNS-construction \cite{Ha},
any such functional gives rise to a representation $\po$ of
$\Al$ on some Hilbert space $\Ho$, and there exists a cyclic
vector\footnote{This means that $\po ( \Al ) {\Omega}_{\omega}$
is dense in $\Ho$ } $\Ooo \in \Ho$ such that
\begin{equation} \omega (A) \ = \ ( \Ooo, \po (A) \Ooo ), \ A \in \Al.
\end{equation}
We note that according to this view the vector states in the original
Hilbert space $\Hh$ induce the identical representation of $\Al$.
Since this representation is distinguished by the existence of a vacuum
state, we call it vacuum representation. It may be regarded as the
defining (faithful) representation of the theory, in accord with
common practice in the actual construction of field theoretic
models based on vacuum functionals.

The states of physical interest correspond to functionals $\omega$ on
$\Al$ which are locally normal with respect to the vacuum representation
\cite[Sec.\ III.3.1]{Ha}. We recall that this property implies,
in view of the Reeh-Schlieder-Theorem, that the
restrictions of these functionals to any local algebra,
$\omega | \, \AO$, can be represented
by vectors $\Omega_{\Oo}$ in the vacuum Hilbert space
$\Hh$ \cite[Sec.\ V.2.2]{Ha},
\begin{equation} \omega ( A ) = ( \Omega_{\Oo}, A \Omega_{\Oo} ), \quad
A \in \AO. \end{equation}
This fact is of interest in the present investigation since it shows
that for the short distance analysis of physical states it suffices
to consider the vector states in the vacuum representation.

We turn now to the analysis of the momentum space properties of the
operators $A \in \Al$. To this end we consider the Fourier
transforms of the operator valued functions $x \rightarrow
\ax ( A ), \; x \in \RR^4$, which are defined in the sense of
distributions. In fact, for each $f \in L^1 ( \RR^4 )$ the expression
\begin{equation} \af ( A ) \doteq
\int dx \, f ( x ) \ax ( A ) \end{equation}
exists as a Bochner integral in $\Al$ because of the norm continuity of
$x \rightarrow \ax ( A )$, and $|| \af ( A )|| \leq
||f||_1 ||A||$.

\vspace{0.5em}
\nin {\em Definition:\/} Let $A \in \Al$. The support of $A$
in momentum space is the smallest closed subset $\OT \subset \RR^4$
such that $\af ( A ) = 0$ for all $f \in L^1 ( \RR^4 )$ with
$\mbox{supp} \fT \subset \RR^4 \backslash \OT$, where $\fT$ denotes the
Fourier transform of $f$. (In more physical terms, $\OT$ may be
called the energy-momentum
transfer of $A$.) The subspace of all operators $A \in \Al$
with support in momentum space in a given set $\OT \subset \RR^4$ is
denoted by $\AOT $.

\vspace{0.5em}
The condition of locality imposes strong restrictions on the support
properties of local operators $A$ in momentum space.
Let us first remark that the support cannot be a compact
set, because otherwise the functions $x \rightarrow \ax ( A )$
would be entire analytic. This in turn would imply that the commutator
functions $x \rightarrow [ \ax ( A ) , B ] $, being $0$ for
any local operator $B \in \Al$ and sufficiently large spacelike
$x$, would
vanish for all $x$. In view of the irreducibility of $\Al$, this
is only possible if $A$ is a multiple of the identity.
By a similar argument it also follows that the support cannot be
contained in any salient cone of $\RR^4$.

It does not seem simple to specify in more explicit terms the
possible support properties of (bounded) local operators in
momentum space. But we  need no such detailed information.
What is of interest here is the fact that there exist in each local
algebra $\AO$ operators whose support in momentum space is
all of $\RR^4$. As a matter of fact this case is generic in a certain
sense, as will become clear from the subsequent argument.
\begin{Lemma}
Suppose that the local algebras are non-trivial for
every open spacetime region, $\AO \neq \CC \cdot 1$. Then each of these
algebras contains operators with support $\RR^4$ in momentum space.
\end{Lemma}
\nin {\em Proof:\/} The proof of this result is based on two well known
facts, cf. for example \cite[Sec.\ VII]{Bo2}. First, for any local
operator $A \in \AOi{0}, \, A \neq c \, 1$, the Fourier
transform of $x \rightarrow \ax ( A ) \Omega$, where $\Omega$ is the
vacuum vector, has Lorentz invariant support which contains
a hyperboloid $H_\mu = \{ p \in \RR^4 \, : p^2 = \mu^2,  p_0 \geq  0 \}$
for some $\mu \geq 0$. Second, if $f,g \in L^1 ( \RR^4 )$ are such
that $\af ( A ) \Omega$ and $\ag ( A ) \Omega$ are different from
$0$, then the function $x \rightarrow \af ( A ) \ax ( \ag ( A ) )
\Omega$ does not vanish for large spacelike $x$, as may be seen from the
cluster theorem. Being the boundary value of
an analytic function in the forward tube $\RR^4 + i V_+$,
because of the spectrum condition, it can therefore not vanish on
open subsets of $\RR^4$.

Now let $q_i, i \in \NN$, be a countable dense subset of points
on the hyperboloid $H_\mu$
and let $\varepsilon_k, k \in \NN$, be a sequence of positive numbers
tending to $0$. For $i,j,k \in \NN$ we define open neighbourhoods of
the points $( q_i + q_j )$,
$$ \Delta_{ijk} \doteq \{ p \in \RR^4 : | p - q_i - q_j | <
\varepsilon_k \}$$
and consider the corresponding sets
$$ N_{ijk} = \{ x \in \RR^4 : E ( \Delta_{ijk} ) A \ax ( A ) \Omega
= 0 \},$$
where $E (\, . \,) $  is the spectral resolution of the
translations $U \,|\,  \RR^4$.
Making use of the spectrum condition it follows that $N_{ijk}$
either does not contain any open set or coincides with all of $\RR^4$.
The latter possibility can be excluded since it would imply that
$ E ( \Delta_{ijk} ) \alpha_{x'} ( A ) \alpha_{x''} ( A ) \Omega
= 0 $ for all $ x', x'' \in \RR^4$. From this it would in turn
follow that for any choice of functions $f_i, f_j \in L^1 ( \RR^4 )$
whose Fourier transforms have support in a ball of radius
$\varepsilon_k / 3$ about $q_i$ and $q_j$, respectively, there holds
$$ \alpha_{f_i} ( A ) \ax ( \alpha_{f_j} ( A ) ) \Omega =
E ( \Delta_{ijk} ) \alpha_{f_i} ( A ) \ax ( \alpha_{f_j} ( A ) ) \Omega
 =  0 $$ for all $x \in \RR^4$. But this is incompatible with the
facts mentioned above. Hence each $N_{ijk}$ is a closed nowhere
dense subset of $\RR^4$, or empty. Since the
countable union of such sets is meager (of first category) its
complement is dense in $\RR^4$ and there exists in each
neighbourhood of the origin in $\RR^4$ a point $y$ which is not
contained in any one of the sets $ N_{ijk} $. Hence, setting
$B = A \ay ( A )$, there holds
$$ E ( \Delta_{ijk}) B  \Omega  \neq 0 \quad  \mbox{for all} \quad
i,j,k \in \NN.$$

\nin Next we consider the sets
$$ N_{ijk,lmn} = \{ x \in \RR^4 : E ( \Delta_{ijk} ) B \ax (B^*)
E ( \Delta_{lmn} ) = 0 \}. $$
Multiplying the operator in the defining condition of this
set from the right by $U(x)$ we see, using once more the spectrum
condition, that also this set cannot contain any open subset,
unless it coincides with $\RR^4$. The latter possibility can again be
excluded since it would imply, by an application of the mean
ergodic theorem to the unitary group $U \,|\, \RR^4$, that
$ E ( \Delta_{ijk} ) B E ( \{ 0 \} ) B^* E ( \Delta_{lmn} ) = 0 $,
which would be in conflict with the properties of $B$ established
before.
Hence also the sets $N_{ijk,lmn}$ are closed and nowhere dense. So
again there exist points $z$ in any neighbourhood of the origin
which are not contained in any one of these sets,
$$ E ( \Delta_{ijk} ) B \az (B^*)  E ( \Delta_{lmn} ) \neq 0 \quad
\mbox{for all} \quad i,j,k,l,m,n \in \NN.$$
It follows from this result that the operator
$ C = B \az ( B^* )$ has support in momentum space which contains
all points $(q_i + q_j - q_l - q_m)$. For if $f \in L^1 ( \RR^4 )$ is
a function
whose Fourier transform $\fT$ is equal to $1$ in any given
neighbourhood of such a point and equal to $0$ in the complement
of any slightly larger region there holds
$$ E ( \Delta_{ijk}) \af ( C ) E ( \Delta_{lmn} ) =
E ( \Delta_{ijk}) C  E ( \Delta_{lmn} ) \neq 0$$
for sufficiently large $k,n$, and consequently $\af ( C ) \neq 0$.
Since the set of points $(q_i + q_j - q_l - q_m)$
is dense in $\RR^4$ this implies that $C$ has support $\RR^4$.
We finally recall that the points $y, z$ in the construction of $C$
from the original operator $A \in \AOi{0}$ can be chosen
as close to $0$ as one wishes. Hence there exists
for any region $\Oo$ which contains the closure of
$\Oo_0$ in its interiour an
operator $C \in \AO$ with the stated support properties in momentum
space. $\Box$

\vspace{6pt}
Whereas local operators cannot have, in the strict mathematical sense,
compact support in momentum space,
it is an important consequence of their continuity properties
with respect to translations that they have ``almost compact support",
as will be explained now.
\begin{Lemma}
 Let the local algebras be non-trivial
and let $\Oo \subset \RR^4$ be given. \\
\nin (i) For any $A \in \AO$ and $\varepsilon > 0$ there exists a
closed ball $\OT \subset \RR^4$ centered at $0$ and an operator
$\AaT \in \AT ( \OT )$ such that
$|| A - \AaT || \leq \varepsilon || A ||$. \\
\nin (ii) Conversely, given any such ball $\OT$
there exist
an operator $A \in \AO$ and operators $ \AaT _\mu, \mu \geq
1$, with support $\mu \OT$ in momentum space such that
$|| A - \AaT _\mu || \leq \mu^{-1} ||A||$.
\end{Lemma}
\nin {\em Proof:\/} (i) Let $f \in {\cal S} ( \RR^4 )$ be any
test function
whose Fourier transform has support in some ball $\OT$ about
$0$, does not
vanish in its interiour, and satisfies $\int dx \, f(x) = 1$. Setting
$f_\mu ( x ) \doteq \mu^4 f ( \mu x ), \mu \geq 1$, we define
operators $\AaT _\mu \doteq \alpha_{f_\mu} ( A )$ which are,
by construction, elements of $\AT ( \mu \OT )$. Moreover, there
holds $|| A - \AaT _\mu || \leq \int dx \, | f ( x ) | \,
|| \alpha_{\mu^{-1} x} ( A ) - A ||$. In view of the continuity
properties of $A$ with respect to translations, the right hand side
of this inequality can be made arbitrarily small for sufficiently large
$\mu$.

(ii) According to the preceding lemma there exist local operators
$A \in \Al ( \Oo )$ which have support $\RR^4$ in momentum space.
Moreover, there also exist such operators for which the corresponding
operator functions $x \rightarrow \ax ( A )$ are differentiable in norm.
(Since all of these functions are norm continuous, the
differentiability can be accomplished by convolution with suitable test
functions whose Fourier transforms do not vanish.)  We pick any
such operator $A$ and consider the corresponding operators
$\AaT _\mu$  introduced in the preceding step. Because of the
support properties of $A$ in momentum space the operators
$\AaT _\mu$ have support $\mu \OT$ and because of
the differentiability of $ x \rightarrow \ax ( A ) $ there holds
$|| \ax ( A ) - A || \leq c_A \, | x |, \, x \in \RR^4 $. Hence it
follows from the estimate given in the preceding step that
$|| A - \AaT _\mu || \leq \mu^{-1} c_{A,f}$,
               where $c_{A,f}$ is a positive
constant depending on $A$ and $f$.
If $c_{A,f} \leq || A ||$ the statement follows.
Otherwise we consider the operator $A' \doteq \zeta 1 + \eta A$, where
$\eta =
{c_{A,f}}^{-1} || A ||$ and the number $\zeta$ is chosen in such a
way that $ || A' || = || A || $. This operator is still an element of
$\AO$ and the corresponding operators $\AaT _\mu ' \doteq \zeta 1 +
\eta \AaT _\mu$ still have support $\mu \OT$ in momentum space since a
non-zero
c-number factor does not change the support properties of an operator
and the unit operator has support $\{0\}$. It thus follows from the
preceding estimate that $ || A' - \AaT _\mu ' ||
                                             \leq \mu^{-1} || A' || $,
which completes the proof. $\Box$

\vspace{6pt}
Although the size of the regions $ \Oo $,
$ \OT $ in the second part of this statement is in principle
arbitrary, it is clear that the respective operators $A$ will be close
in norm to the identity (i.e., the constant $\eta$ in the last step of
the above argument will be very small) if the regions do not comply
with the uncertainty principle. In order to obtain operators which
do not have a dominant c-number part $\zeta 1$, the product of the
diameters of $\Oo$ and $\OT$ should be of the order of Plancks
constant. It is conceivable that in some theories fitting into our
general setting the latter condition does not suffice to ensure
the existence of
non-trivial (``quantum") operators for arbitrarily small regions $\Oo$.
The orbit of any local operator under the action of the renormalization
group would then tend in norm to a multiple of the identity,
hence the quantum correlations between ``comparable observables" would
disappear in the scaling limit. This possibility will be discussed in
Sec.\ 4 in more detail.
\section{Scaling algebra and renormalization group}
\setcounter{equation}{0}
\noindent
In this section we introduce the concept of scaling algebra
which allows us to express the basic ideas of the renormalization
group in the algebraic setting. Our starting point is a local,
Poincar\'e covariant net
$\Al, \aP$
as described in Sec.\ 2, which is assumed to be given. It is called
the underlying net or underlying theory.

To fix ideas we assume that the net
$\Al, \aP$
is defined at spatio-temporal scale
$\lambda = 1$.
As was discussed in the Introduction, one then obtains the
corresponding nets at any other scale
$\lambda >0$
by setting
\begin{equation}
\Oo \to \AlO \doteq \Al(\lambda\Oo)\,.
\end{equation}
The Poincar\'e transformations at scale
$\lambda$
are given by
\begin{equation}
\aLx^{(\lambda)} \doteq \alpha_{\Lambda,\lambda x}\,,
   \quad   \Lx \in \Pg \,.
\end{equation}
Note that
$\Al_{\lambda},\alpha^{(\lambda)}_{\Pg}$
defines again a local, Poincar\'e covariant net over Minkowski
space which complies with the conditions given in Sec.\ 2. Thus,
in accord with the procedure in the conventional field
theoretical setting, we keep Minkowski space fixed and interpret
the properties of the underlying theory at small scales
$\lambda$
in terms of the modified theories (nets)
$\Al_{\lambda},\alpha^{(\la)}_{\Pg}$.

It is apparent that the nets
$\Al_{\lambda},\alpha^{(\la)}_{\Pg}$
describe, for different values of
$\lambda$,
in general distinct theories (with different energy-momentum
spectrum, collision cross sections, ``running'' coupling constants
etc.). Within the algebraic setting these differences find there
formal expression in the fact that the corresponding nets are
non-isomorphic. Conversely, any two local, Poincar\'e-covariant
nets which are isomorphic are physically indistinguishable and
consequently represent the same theory. We recall the notion of
net isomorphism in the following definition.
\\[6pt]
{\em Definition:\/} For $j = 1, 2$, let
$\Oo \to \Al^{(j)}(\Oo)$, $\alpha^{(j)}_{\Pg}$
be two local, Poincar\'e covariant nets on Minkowski space with
$C^*$-inductive limits
$\Al^{(j)}$.
The two nets are said to be {\em isomorphic}
if there is an
isomorphism
$\phi : \Al^{(1)} \to \Al^{(2)}$
which preserves localization,
$$ \phi(\Al^{(1)}(\Oo)) = \Al^{(2)}(\Oo)\, , \quad \Oo \subset \RR^4 $$
and intertwines the Poincar\'e transformations,
$$ \phi \ckl \alpha^{(1)}_{\Lambda,x} =
   \alpha^{(2)}_{\Lambda,x} \ckl \phi \,,\quad \Lx \in \Pg\,.
    $$
Any such isomorphism
$\phi$
is called a {\em net isomorphism.} A net isomorphism which maps a
given net onto itself is called an {\em internal symmetry.}

\vspace{6pt}
According to this definition the nets
$\Al_{\lambda},\alpha^{(\la)}_{\Pg}$
are isomorphic for different values of
$\lambda$
if and only if dilations are a (geometrical)
                               symmetry of the underlying theory.
The physical content of the theory is then invariant under changes of
the spatio-temporal scale.
We are here primarily interested in those cases where the underlying
theory does {\em not} possess such a symmetry. Consequently we cannot
rely on the notion of net-isomorphism in order to compare the
properties of the theory at different scales.

                                              This comparison can be
accomplished, however, with the help of the renormalization
group transformations
$\Rl$,
considered in the Introduction, which map the underlying net
$\Al$
onto the nets
$\Al_{\lambda}$
at any other scale (though without preserving algebraic relations).
As was explained, the transformations
$\Rl$
are used in the conventional setting of quantum field theory to
identify particular observables at different scales. Yet such a
detailed information is not necessary for the interpretation of
the short distance properties of a theory. All what matters is
that renormalization group transformations do not change the
fundamental physical units
$c$ and $\hbar$
and are continuous, cf.\ conditions (i), (ii) and (iii) in the
Introduction. As we shall see, it is not even necessary for the
short distance analysis to specify the transformations
$\Rl$
explicitly. It suffices to have control on the phase space
properties of the orbits
$\lambda \to \Rl(A)$, $\la > 0$,
of local observables
$A \in \Al$
under these transformations. Thus, instead of starting from some
{\em ad hoc} choice of the transformation
$\Rl$,
we will consider the set of all operator functions of the
scaling parameter
$\la$
with values in
$\Al$
which exhibit the relevant features of the orbits of observables
under the action of the renormalization group. These functions
will constitute the elements of the scaling algebra.

Before we can turn to the actual definition of this algebra we have
to elaborate the relevant properties of the functions
$\la \to \Rl(A)$, $A \in \Al$.
Given any
$A \in \AO$
it follows from condition (i) in the Introduction that
\begin{equation}
 \Rl(A) \in \Al_{\la}(\Oo) = \Al(\la \Oo)\,, \quad \la > 0\,.
\end{equation}
Thus these functions have specific localization properties in
configuration space. Next, let us determine the properties of
$ \Rl(A) $
in momentum space. As was explained in Sec.\ 2, local operators
$A \in \AO$
cannot have compact support in momentum space, but they have almost
compact support in the sense of part (i) of Lemma 2.2. More
precisely, given
$\delta > 0$,
there is a compact set
$\OT \subset \RR^4$
and an operator
$\widetilde{A} \in \AOT$
such that
$||\, A - \widetilde{A} \,|| < \delta \, ||\,A\,||$.
According to condition (ii) there holds
$\Rl(\widetilde{A}) \in \AT_{\la}(\OT) = \AT(\la^{-1}\OT)$.
Thus, making use of the continuity properties of
$\Rl$,
cf.\ condition (iii), we conclude (choosing
$\delta$
sufficiently small) that for any
$\varepsilon > 0$
there is some compact set
$\OT$
such that
\begin{equation}
\Rl(A) \in \AT(\la^{-1}\OT) + \varepsilon\,{\frak U}\,, \quad \la > 0
\end{equation}
where $\frak U$ is the unit ball in
$\Al$.
Hence in this approximate sense the orbits of local observables
under renormalization group transformations have characteristic
support properties in momentum space as well. Roughly speaking, the
volume of phase space occupied by these operators is independent
of the scale
$\la$.

It will greatly simplify our discussion that the mildly
cumbersome relation (3.4) can be replaced by a continuity
condition with respect to space-time translations, as is
shown in the subsequent lemma.
\begin{Lemma}  Let
$\la \to A_{\la}$, $\la >0$,
be a function with values in
$\Al$
which is uniformly bounded in norm. Then the following two
     statements are
equivalent.   \\
(i) For each
     $\varepsilon > 0$
     there exists a compact set
     $\OT \subset \RR^4$
     such that
     $A_{\la} \in \AT(\la^{-1}\OT) + \varepsilon\, {\frak U}$
     for all
     $\la > 0$.   \\
(ii) $\sup_{\la >0}\,||\,\alpha_{\la x}
                                        (A_{\la}) - A_{\la}
      \, || \to 0 $
     for
     $x \to 0$.
\end{Lemma}
{\em Proof:\/} That the first statement implies the second one may be
seen as follows. Let
$\varepsilon > 0$
and let
$\widetilde{A}_{\la} \in \AT(\la^{-1}\OT)$
be such that
$||\,A_{\la} - \widetilde{A}_{\la} \,|| \leq \varepsilon$,
$\la >0$, and consequently
$$ \sup_{\la > 0}\,||\,\alpha_{\la x}(A_{\la}) -A_{\la}\,||
\leq \sup_{\la > 0} ||\, \alpha_{\la x}(\widetilde{A}_{\la}
 ) - \widetilde{A}_{\la} \,|| + 2\varepsilon \,.
 $$
Since
$A_{\la}$
is uniformly bounded for
$\la > 0$
the same holds true for
$\widetilde{A}_{\la}$.
We pick any function
$f \in L^1(\RR^4)$
whose Fourier transform is equal to $1$ in a neighbourhood of
$\OT$
and set
$f_{\la}(x) \doteq \la^{-4}f(\la^{-1}x)$.
Because of the support properties of
$\widetilde{A}_{\la}$
in momentum space we have
$\widetilde{A}_{\la} = \alpha_{f_{\la}}(\widetilde{A}_{\la})$,
hence
$$ \sup_{\la}\,||\,\alpha_{\la x}(\widetilde{A}_{\la})
 -\widetilde{A}_{\la} \,|| \leq
  \int dy\,|f(y-x)-f(y)|\cdot \sup_{\la >0}\,
  ||\,\widetilde{A}_{\la}\,|| \,.$$
Since the right hand side of this inequality vanishes for
$x \to 0$,
the second statement follows.

For the proof of the converse direction we put
$\eta(x) \doteq \sup_{\la >0}\,||\,\alpha_{\la x}(A_{\la}
      )    -  A_{\la}\,|| $.
By assumption this function is bounded and vanishes for
$x \to 0$.
Given
$\varepsilon > 0$,
we can therefore find some function
$g \in L^1(\RR^4)$
such that
$\int dx\,g(x) = 1$, $\int dx\,|g(x)|\eta(x) \leq \varepsilon$
and the Fourier transform of $g$ has compact support in some
(sufficiently large) region
$\OT$.
Proceeding to the scaled functions
$g_{\la}$
as above it follows that the operators
$\widetilde{A}_{\la} \doteq \alpha_{g_{\la}}(A_{\la})$
have support in momentum space in the region
$\la^{-1}\OT$.
Since
$\int dx\,g_{\la}(x) = 1$
there holds also
$||\,A_{\la} - \widetilde{A}_{\la} \,|| \leq
 \int dx\,|g(x)|\eta(x) \leq \varepsilon $ for $\la > 0$,
proving the first statement.  $\Box$
\\[6pt]
In view of this lemma and the fact that the functions
$\la \to \Rl(A)$
are uniformly bounded, cf.\ condition (iii) in the Introduction, we can
proceed from relation (3.4) to the equivalent statement
\begin{equation}
 \sup_{\la >0}\,||\,\alpha_{\la x}(\Rl(A)) - \Rl(A)\,||
 \to 0 \quad {\rm for} \quad x \to 0\,.
\end{equation}
Taking into account that angular momentum has the dimension of
$\hbar$
and that renormalization group transformations do not change
the unit of action one
is led by a similar spectral analysis of the functions
$\la \to \Rl(A)$
with respect to the Lorentz transformations $\alpha_{
{\cal L}^{\uparrow} _{+}}$  to the conclusion that also
\begin{equation}
 \sup_{\la >0}\,||\,\aL(\Rl(A)) -\Rl(A) \,|| \to 0
 \quad {\rm for} \quad \Lambda \to  1,\ \Lambda \in
 {\cal L}^{\uparrow}_{+} \,.
 \end{equation}
We regard the properties (3.3), (3.5) and (3.6) as the distinctive
features of the orbits of local operators under renormalization
group transformations. As a matter of fact one can give arguments
that any function
$\la \to A_{\la}$
with these properties arises from some transformation
which satisfies conditions (i), (ii) and (iii) in the
Introduction with arbitrary precision. We may therefore forget the
underlying transformations and consider the set of {\em all}
functions
$\la \to A_{\la}$
with these properties. Given any such function we interpret its
values
$A_{\la}$
as elements of the respective nets (theories)
$\Al_{\lambda},\alpha^{(\la)}_{\Pg}$.
Thus these functions provide information as to which operators
are identified at different scales. Since the operators
$A_{\la}$
are only restricted by phase space conditions, this view
may appear to be problematic if one thinks of the rigid assignment
of operators at different scales in the conventional approach
to the renormalization group. But let us recall that according
to the algebraic point of view the physical information of a
theory is contained in (and can be recovered from) the net.
It is not necessary to fix the interpretation of individual operators.
Phrased differently: the physically relevant information about the
action of the renormalization group is contained in the particular
embedding of the nets
$\Al_{\lambda},\alpha^{(\la)}_{\Pg}$
at different scales
$\la$
into each other, which is provided by the functions
$\la \to A_{\la}$.
It is {\em not} contained in the individual orbits.

We emphasize that this approach provides a redundant description
of the renormalization group since the orbits induced by all
admissible transformations
$\Rl$
are taken into consideration. This causes some minor technical
complications which can be handled, however, by standard
functional analytic  techniques.
What one gains on the other hand by this method is a general
formalism for the renormalization group analysis which is
canonically associated with any given theory and free of
arbitrariness. Moreover, it circumvents the actual construction
of the renormalization group transformations
$\Rl$.

To stress the latter point let us note that functions
$\la \to A_{\la}$
with the desired properties can easily be constructed in
abundance. This follows from the second part of Lemma 2.2.
According to that statement there exist for fixed regions
$\Oo,\OT$
and any
$\la > 0$
operators
$A_{\la} \in \Al(\la\Oo)$, $||\,A_{\la}\,|| =1$,
and
$\widetilde{A}_{\la,\mu} \in \AT(\mu \la^{-1}\OT)$, $\mu \geq 1$,
such that
$||\,A_{\la} - \widetilde{A}_{\la,\mu}\,|| \leq \mu^{-1}$.
Thus the corresponding functions
$\la \to A_{\la}$
satisfy conditions (3.3) and (3.4), hence also (3.5). In order to
satisfy also condition (3.6) it suffices to average the operators
$A_{\la}$, $\la >0$
over a fixed, arbitrarily small neighbourhood
$\cal N$
of unity in
${\cal L}^{\uparrow}_{+}$,
$\overline{A}_{\la} \doteq \int_{\cal N} d\Lambda\,
 \aL(A_{\la})$,
where
$d\Lambda$
is the Haar-measure. The resulting functions
$\la \to \overline{A}_{\la}$
then have all desired properties.

After this motivation of our particular approach to the
renormalization group analysis let us turn now to the definition
of the scaling algebras. As was explained, we consider functions
$\Au : \RR^+ \to \Al$
from the domain
$\RR^+$
of the scaling parameter
$\la$
to the underlying algebra of observables. (In order to distinguish
these functions from elements of
$\Al$
we mark them in the following by underlining.) Since we want to
interpret the values
$\Aul$
of these functions as elements of the local, covariant nets
$\Al_{\la},\alpha^{(\la)}_{\Pg}$,
it is natural to induce the following algebraic structures:
given any two functions
$\Au,\Bu$
and
$a,b \in \CC$
we set for $\la > 0$
\begin{eqnarray}
 (a\Au + b \Bu)_{\la} & \doteq & a\Aul + b\Bu_{\la} \nonumber \\
 (\Au \cdot \Bu)_{\la} & \doteq& \Aul \cdot \Bu_{\la} \\
 (\Au^*)_{\la} & \doteq & \Aul^* \,. \nonumber
\end{eqnarray}
Thereby the functions
$\Au$
acquire the structure of a unital *-algebra, the unit being given
by
$\underline{ 1}_{\la} =  1$.
Moreover, since we are only interested in uniformly bounded
functions it is natural to introduce the norm
\begin{equation}
||\,\Au\,|| \doteq \sup_{\la > 0}\,||\,\Aul\,||
\end{equation}
which in fact is a $C^*$-norm. Bearing in mind that
$\alpha^{(\la)}_{\Lambda,x} = \alpha_{\Lambda,\la x}$,
the induced action of the Poincar\'e transformations on the
functions is given by
\begin{equation}
(\aLxu(\Au))_{\la} \doteq \alpha_{\Lambda,\la x}(\Aul)\,.
\end{equation}
It follows that the continuity requirements (3.5) and (3.6) can
be expressed in the simple form
\begin{equation}
||\,\aLxu(\Au) -\Au\,|| \to 0 \quad {\rm for}
\quad \Lx \to ( 1,0)\,.
\end{equation}
It remains to impose on the functions the localization condition
(3.3). This is accomplished with the help of the following
definition.
\\[6pt]
{\em Definition:\/} Let
$\Oo \subset \RR^4$
be any open, bounded region. Then
$\Alu(\Oo)$
denotes the set of all uniformly bounded functions
$\Au$
which are continuous with respect to Poincar\'e transformations
in the sense of relation (3.10) and satisfy
$$ \Aul \in \Al(\la\Oo)\,,\quad \la > 0\,. $$

\vspace{6pt}
Since each
$\Al(\la \Oo)$
is a $C^*$-algebra it follows that
$\Alu(\Oo)$
is a $C^*$-algebra as well: it is stable under the algebraic
operations (3.7) and complete with respect to the $C^*$-norm (3.8).
(Note that
$||\,\aLxu(\Au)\,|| = ||\,\Au\,||$,
hence the limit of any Cauchy sequence whose elements satisfy
the continuity condition (3.10) again satisfies this condition.)
It is also apparent from the definition that
$\Alu(\Oo)$
is monotonous with respect to
$\Oo$,
\begin{equation}
\Alu(\Oo_1) \subset \Alu(\Oo_2) \quad {\rm if}
\quad \Oo_1 \subset \Oo_2\,.
\end{equation}
Thus the assignment
$\Oo \to \Alu(\Oo)$
defines a net of $C^*$-algebras over Minkowski space.
Since
$\Oo_1 \subset \Oo_2'$
implies that also
$\la \Oo_1 \subset (\la \Oo_2)'$, $\la > 0$
(the velocity of light, $c$, is kept constant at each scale) it
follows from the locality of the underlying theory that the net
$\Alu$
is local, too,
\begin{equation}
\Au \,\Bu = \Bu\,\Au \quad {\rm for} \quad
\Au \in \Alu(\Oo_1), \Bu \in \Alu(\Oo_2) \quad
{\rm if} \quad \Oo_1 \subset \Oo_2' \,.
\end{equation}
Finally, because of the  equality of sets
$(\Lambda  \la \Oo + \la x) = \la(\Lambda\Oo + x)$, $\la>0$,
it follows from (3.9) and covariance of the underlying theory that
\begin{equation}
\aLxu(\Alu (\Oo)) = \Alu(\Lambda\Oo + x)\,.
\end{equation}
Hence the Poincar\'e transformations
$\auP$
are automorphisms of the net
$\Alu$.
We have thus arrived at a mathematical setting which is similar
to our starting point.
\\[6pt]
{\em Definition:\/} The local, covariant net
$\Alu,\auP$
is called {\em scaling net} of the underlying theory. The
                     $C^*$-inductive
limit
of all local algebras
$\Alu(\Oo)$
is called {\em scaling algebra} and denoted by
$\Alu$.
(As is common practice, we denote the net and the corresponding
global $C^*$-algebra by the same symbol.)
\\[6pt]
{\em Remark:\/} The elements of the global scaling algebra
$\Alu$
can still be represented by certain specific functions
$\la \to \Aul$
with values in
$\Al$.
\\[6pt]
It is straightforward to describe in this formalism a change of
the spatio-temporal scale. Such changes are induced by an
automorphic action
$\underline{\sigma}_{\RRf^{ +}}$
of the multiplicative group
$\RR^+$
on the scaling algebra
$\Alu$,
given for any
$\mu \in \RR^+$
by
\begin{equation}
(\smu(\Au))_{\la} \doteq \Au_{\mu\la}\,, \quad \la > 0\,.
\end{equation}
As is easily verified, there holds
\begin{equation}
\smu(\Alu(\Oo)) = \Alu(\mu\Oo)\,, \quad \Oo \subset \RR^4
\end{equation}
and
\begin{equation}
\smu \ckl \aLxu = \underline{\alpha}_{\Lambda,\mu x}
\ckl \smu\,, \quad \Lx \in \Pg\,.
\end{equation}
According to our previous discussion the scaling
transformations
$\underline{\sigma}_{\RRf^{ +}}$
are to be viewed as renormalization group transformations
which relate the observables at different scales. They appear
in the present setting as geometrical symmetries; this greatly
simplifies their analysis.

The scaling algebra
$\Alu$
comprises the orbits of local observables under all possible
renormalization group transformations which comply with the
basic conditions given in the Introduction. So in this sense
it is maximal. In applications of the formalism it may sometimes
be convenient to impose further constraints on the orbits which
amounts to proceeding to subnets of
$\Alu$.
For example, it seems natural to restrict attention to those
elements of
$\Alu$
on which the scaling transformations
$\smu$, $\mu > 0$,
act norm continuously. Let us note, however, that continuity
at
$\mu =0$
(from the right) would be too strong a requirement. For only
multiples of the identity
$\underline{1}$
have this property. In the present investigation we will not
impose any further constraints on the functions
$\Au$
and work with the maximal net
$\Alu$.

It is our objective to study the properties of the physical states
of the underlying theory at arbitrarily small scales. To this
end we will lift these states to the scaling algebra and study
their behaviour under scaling transformations. Before we explain
this procedure we introduce some notation. Let
$\ou$
be any state on the scaling algebra
$\Alu$
and let
$(\pu,\Hu,\Ou)$
be the corresponding GNS-representation. The kernel of
$\pu$
(i.e., the closed two-sided ideal consisting of all elements
$\Au \in \Alu$
for which
$\pu(\Au) = 0$)
will be denoted by
${\rm ker}(\pu)$
and the quotient of
$\Alu$
with respect to
${\rm ker}(\pu)$
by
\begin{equation}
\Alu^{\pu} \doteq \Alu/{\rm ker}(\pu) \simeq \pu(\Alu)\,.
\end{equation}
Here the symbol
$\simeq$
indicates the well-known fact \cite[Cor. 1.8.3]{Di} that the
respective $C^*$-algebras are isomorphic. The canonical
isomorphism
$\psi$
between these algebras is given by
$\psi(\Au^{\pu}) = \pu(\Au)$,
where
$\Au^{\pu}$ is the class of
$\Au \in \Alu$
modulo
${\ker}(\pu)$.
The projection of
$\ou$
to the quotient
$\Alu^{\pu}$
is denoted by
${\rm proj}\,\ou$
and given by
\begin{equation}
{\rm proj}\,\ou \doteq (\Ou,\psi(\,.\,)\Ou)\,.
\end{equation}
The physical interpretation of the states
$\ou$
will be based on their projections
${\rm proj}\,\ou$,
regarded as states on the net
\begin{equation}
\Oo \to \Alu^{\pu}(\Oo) \doteq \Alu(\Oo)/{\rm ker}(\pu)
\end{equation}
on Minkowski space. These nets are again local. Moreover, if
${\rm ker}(\pu)$
is invariant under the Poincar\'e transformations
$\auP$
one can also define an automorphic action of the Poincar\'e
group on
$\Al^{\pu}$,
setting for
$\Lx \in \Pg$
\begin{equation}
\alpha^{\pu}_{\Lambda,x}(\Au^{\pu}) \doteq
(\aLxu(\Au))^{\pu} \,,\quad \Au^{\pu} \in \Alu^{\pu}\,.
\end{equation}
In this way any suitable state
$\ou$
on
$\Alu$
determines a local, covariant net
$\Alu^{\pu},\alpha^{\pu}_{\Pg}$
and a distinguished state
${\rm proj}\,\ou$
on
$\Alu^{\pu}$.

Given a state
$\omega$
on the underlying net
$\Al,\alpha_{\Pg}$
we intend to define its canonical lift
$\ou$
to the scaling algebra
$\Alu$
in such a way that we can recover from it by the above procedure
the underlying net and state. This leads us to the following
definition.
\\[6pt]
{\em Definition:\/} Let
$\omega$
be a state on the underlying global algebra
$\Al$.
Its {\em canonical lift}
$\ou$
on the scaling algebra
$\Alu$
is defined by
$$   \ou(\Au) \doteq \omega(\Au_{\la =1})\,, \quad \Au \in \Alu\,.$$

\vspace{6pt}
\noindent
It is apparent that
$\ou$
is a state on
$\Alu$
and we shall see with the help of the subsequent lemma that it
provides the desired information.
\begin{Lemma} Let
$\omega$
be a state on
$\Al$, let
$(\pi,{\cal H},\Omega)$ be
the corresponding GNS-representation and let
$\ou$
be the canonical lift of
$\omega$
on
$\Alu$
with corresponding GNS-representation
$(\pu,\Hu,\Ou)$.
There exists a unitary
$W : \Hu \to \Hh$
such that
$W\Ou = \Omega$
and
$$ W\pu(\Au)W^{-1} = \pi(\Au_{\la =1})\,, \quad \Au \in \Alu\,.$$
Moreover, for each bounded region
$\Oo$
it holds that
$$ W\pu(\Alu(\Oo))W^{-1} = \pi(\AO)\,, $$
i.e., $W$ induces an isomorphism between the $C^*$-algebras
$\pu(\Alu)$
and
$\pi(\Al)$
which preserves localization.
\end{Lemma}
{\em Proof:\/} It follows from the definition of the canonical lift
that
$$ (\Ou,\pu(\Au)\Ou) = (\Omega,\pi(\Au_{\la =1})\Omega)\,,
                          \quad \Au \in \Alu\,.  $$
Hence the map
$W : \Hu \to \Hh$
given by
$$ W\pu(\Au)\Ou \doteq \pi(\Au_{\la = 1})\Omega\,,\quad \Au \in \Alu$$
is densely defined, isometric and linear. Setting
$\Au = \underline{1}$
we have in particular
$ W \Ou = \Omega$.
In order to show that the range of $W$ is dense in
$\Hh$
we proceed as follows. Given any region
$\Oo$
and any operator
$A \in \AO$
we define a function
$\Au$,
setting
$\Aul \doteq A$
for
$\la = 1$
and
$\Aul \doteq 0$
for
$\la \neq 1$.
One easily checks that
$\Au \in \Alu(\Oo)$
and consequently
$\pi(\AO) \subset \pi(\Alu_{\la =1}(\Oo))$,
in an obvious notation. The converse inclusion holds by construction
of
$\Alu(\Oo)$,
hence
\begin{equation}
\pi(\Alu_{\lambda = 1}(\Oo)) = \pi(\AO)\,.
\end{equation}
Proceding  to the inductive limit
$\Oo \nearrow \RR^4$
we see that also
$\pi(\Alu_{\la =1}) = \pi(\Al)$.
This shows that the range of $W$ contains the vectors
$\pi(\Al)\Omega$
which are dense in
$\Hh$
according to the GNS-construction. Thus $W$ maps
$\Hu$
onto
$\Hh$
and hence is invertible. Making use of the defining relation of $W$
we conclude that
$$ W\pu(\Au)W^{-1} = \pi(\Au_{\la =1})\,,\quad \Au \in \Alu\,. $$
Hence
$W\, \cdot\, W^{-1}$
defines an isomorphism mapping
$\pu(\Alu)$
onto
$\pi(\Al)$
which, because of relation (3.21), preserves localization.
$\Box$
\\[6pt]
This lemma has several interesting consequences which we list
below. First, let us recall the following standard terminology.
\\[6pt]
{\em Definition:\/} Let
${\frak B}$
be a $C^*$-algebra and let
$\beta$
be a representation of the Poincar\'e group
$\Pg$
by automorphisms of
${\frak B}$.
A representation
$(\pi,\Hh)$
of
${\frak B}$
is said to be covariant with respect to the action of
$\beta_{\Pg}$
(briefly: a {\em covariant representation} of
${\frak B},\beta_{\Pg}$)
if there is
a continuous unitary (possibly projective) representation $V$ of
$\Pg$ on
 $\Hh$ such that for
$\Lx \in \Pg$
$$ V\Lx\pi(B)V\Lx^{-1} = \pi(\beta_{\Lambda,x}(B))\,,\quad B\in
  {\frak B}\,.$$
A covariant representation
$(\pi,\Hh)$
is said to be a {\em vacuum representation} if
$V\,|\,\RR^4$
satisfies the relativistic spectrum condition, $sp\,V\,|\, \RR^4
\subset \overline{V}_{+}$, and there is a
vector
$\Omega \in \Hh$
which is cyclic for
$\pi({\frak B})$
and invariant under the action of
$V\Lx$, $\Lx \in \Pg$.
Finally, a state
$\omega$
on
${\frak B}$
is called a {\em vacuum state} if the corresponding
GNS-representation
$(\pi,\Hh,\Omega)$
is a vacuum representation where
$\Omega$
has the just mentioned properties.
${}$
\begin{Corollary} Let
$\omega$
be a state on
$\Al$
and let
$\ou$
be its canonical lift on
$\Alu$.   \\
(i)  $\ou$
      is a pure state iff
      $\omega$
      is pure.  \\
(ii) The GNS-representation
      $(\pu,\Hu,\Ou)$
      of
      $\Alu,\auP$,
      induced by
      $\ou$,
      is covariant iff the GNS-representation
      $(\pi,\Hh,\Omega)$
      of
      $\Al,\alpha_{\Pg}$,
      induced by
      $\omega$,
      is covariant.
      Moreover, if
      $\underline{U},U$
      denote the representations of
      $\Pg$
      on the respective Hilbert spaces
      $\Hu,\Hh$,
      one obtains
      $$ \underline{U}\Lx = W^{-1}U\Lx W\,,\quad \Lx \in \Pg $$
where
      $W : \Hu \to \Hh$
      is the unitary map appearing in the preceding lemma.
         \\
(iii) $\ou$
      is a vacuum state iff
      $\omega$
      is a vacuum state.
\end{Corollary}
{\em Proof:\/} These statements are an elementary consequence of
Lemma 3.2. We confine ourselves to establishing the second one.
Let
$(\pi,\Hh,\Omega)$
be a covariant representation of
$\Al,\alpha_{\Pg}$.
There hold the equalities (cf.\ Lemma 3.2)
\begin{eqnarray*}
\lefteqn{ W\pu(\aLxu(\Au))W^{-1} =
    \pi((\aLxu(\Au))_{\la =1}) } \\
    & &=\ \pi(\aLx(\Au_{\la =1})) =  U\Lx\pi(\Au_{\la=1})
    U\Lx^{-1} \\
    & & =\ U\Lx W\pu(\Au) W^{-1} U\Lx^{-1} \,.
\end{eqnarray*}
Hence
$(\pu,\Hu,\Ou)$
is a covariant representation of
$\Alu,\auP$,
the corresponding unitary representation
$\underline{U}$
of
$\Pg$
being given by
$\underline{U}\Lx = W^{-1}U\Lx W$, $\Lx \in \Pg$.
The proof of the ``only if'' part of the statement is analogous.
$\Box$
\\[6pt]
Let us apply now these results to the vector states
$\omega$
in the defining (identical) vacuum representation of
$\Al,\alpha_{\Pg}$.
Since this representation is covariant, the same is true for the
representations
$(\pu,\Hu,\Ou)$
of
$\Alu,\auP$
induced by the respective canonical lifts
$\ou$.
As a consequence, the kernels
${\rm ker}(\pu)$
are invariant under the action of
$\auP$.
Hence any such vector state
$\omega$
determines, as outlined above, via its lift
$\ou$
on
$\Alu$
a local covariant net
$\Alu^{\pu},\alpha^{\pu}_{\Pg}$
and a state
${\rm proj}\,\ou$.
Recalling that the $C^*$-algebras
$\Alu^{\pu}$
and
$\pu(\Alu)$
are canonically isomorphic, it follows from Lemma 3.2 and its
corollary that the nets
$\Alu^{\pu},\alpha^{\pu}_{\Pg}$
and
$\Al,\alpha_{\Pg}$
are isomorphic, a net isomorphism
$\phi$
being given by
\begin{equation}
\phi(\Au^{\pu}) = \Au_{\la =1}\,, \quad \Au^{\pu} \in \Alu^{\pu}\,.
\end{equation}
Hence the two nets describe the same physics. Moreover, we have
\begin{equation}
{\rm proj}\,\ou \ckl \phi^{-1} = \omega\,,
\end{equation}
hence with the above choice of
$\phi$
one can also recover from
${\rm proj}\,\ou$
the underlying state
$\omega$.

Let us note that any two isomorphisms between the above nets are
related by an internal symmetry transformation of
$\Al,\alpha_{\Pg}$.
Thus even if one forgets about the specific assignment (3.23) one can
recover from
${\rm proj}\,\ou$
the state
$\omega$
up to some internal symmetry transformation.
The apparent loss of information about
$\omega$
is irrelevant since all physical states look alike at very small
scales, as we shall see. Moreover, if the internal
symmetries of
$\Al,\alpha_{\Pg}$ (provided there are any)
 are not spontaneously broken in the underlying vacuum
representation, then the respective vacuum state
is invariant under these symmetries. Hence in this relevant case
one can recover without ambiguities the properties of
the vacuum
from its projected lift.
These facts show that for the physical interpretation of the
states
${\rm proj}\,\ou$
there is no need to rely on the framework of the underlying theory.
This point will be of relevance in our discussion of the
scaling limit, where an ``underlying theory'' has yet to be
defined.

Having seen how the physical interpretation of the states
$\omega$
in the underlying theory can be recovered from their canonical
lifts
$\ou$
it is now straightforward to obtain information about the
properties of these states at any other scale
$\la > 0$.
One simply has to compose the lifted states
$\ou$
with the scaling transformation
$\sglu$,
\begin{equation}
 \ou_{\la} \doteq \ou \ckl \sglu \,.
\end{equation}
Since
$\sglu$
is an automorphism of the scaling algebra
$\Alu$
one can identify the GNS-representation induced by
$\ou_{\la}$
with
$(\pul,\Hu,\Ou)$
where
\begin{equation}
\pul \doteq \pu \ckl \sglu
\end{equation}
and
$(\pu,\Hu,\Ou)$
is the original representation of
$\Alu$
induced by
$\ou$.
It follows from relation (3.16) and the Poincar\'e covariance of
$(\pu,\Hu,\Ou)$
that
${\rm ker}(\pul)$
is invariant under
$\auP$.
Consequently each state
$\ou_{\la}$
determines a local, covariant net
$\Alu^{\pul},\alpha^{\pul}_{\Pg}$
and a state
${\rm proj}\,\ou_{\la}$
on this net. That this net and state have the desired interpretation
is the content of the subsequent proposition.
\begin{Proposition} Let
$\omega$
be a vector state in the defining vacuum representation of
$\Al$,
let
$\ou$
be its canonical lift on
$\Alu$
and let
$\ou_{\la}$, $\la > 0$,
be the corresponding scaled states with GNS-representations
$(\pul,\Hu,\Ou)$.
For each
$\la > 0$
there exists a net isomorphism
$\phi_{\la}$
between the nets
$\Alu^{\pul},\alpha^{\pul}_{\Pg}$
and
$\Al_{\la},\alpha^{(\la)}_{\Pg}$
given by
$$ \phi_{\la}(\Au^{\pul}) \doteq \Aul\,, \quad \Au^{\pul} \in
 \Alu^{\pul}\,. $$
Moreover, there holds
$$ {\rm proj}\,\ou_{\la} \ckl \phi^{-1}_{\la} = \omega $$
where
$\omega$
is regarded as a state on the net
$\Al_{\la},\alpha^{(\la)}_{\Pg}$.
\end{Proposition}
{\em Proof:\/} Let
$\sgl : \Alu^{\pul} \to \Alu^{\pu}$
be the map given by
$$  \sgl(\Au^{\pul}) \doteq (\sglu(\Au))^{\pu}\,,
\quad \Au^{\pul} \in \Alu^{\pul}\,. $$
Since
$\sglu$
is an automorphism of
$\Alu$
it follows that
$\sgl$
is an isomorphism. Moreover, one obtains
$$ \sgl(\Alu^{\pul}(\Oo)) = \Alu^{\pu}(\la \Oo)\,.$$
Noticing that
$\phi_{\la} = \phi \ckl \sgl$,
where
$\phi$
is the isomorphism between
$\Alu^{\pu},\alpha^{\pu}_{\Pg}$
and
$\Al,\alpha_{\Pg}$
defined in relation (3.22), and recalling relation (3.16) it is
straightforward to complete the proof of the statement.
$\Box$
\\[6pt]
In view of this result it is obvious how the states
$\ou_{\la}$
on
$\Alu$
can be used to obtain, in the limit
$\la \searrow 0$,
information on the scaling limit of the underlying theory.
The discussion of this issue is the subject of the subsequent
section.
\section{Scaling limit}
\setcounter{equation}{0}
Let us now investigate the behaviour of physical states
$\omega$
on
$\Al$
under renormalization group transformations in the scaling limit
$\lambda \searrow 0$.
Hence we consider the family of states on
$\Alu$
\begin{equation}
\olu \doteq \ou \ckl \sglu\, , \quad \lambda > 0,
\end{equation}
as a net directed towards
$\lambda = 0$.

Before entering into details, let us point out some at first sight
perhaps unexpected mathematical problem appearing in this analysis.
As explained, we want to determine the properties of states at
small scales with the help of the functions
$\Au \in \Alu$.
Now there holds
$\bigcap_{\lambda > 0}\Al(\lambda \Oo)^{-} = \CC \cdot  1$
(cf.\ the proof of Lemma 4.1 below), hence any function
$\Au$
with the property that
$\Aul$
converges in norm for
$\lambda \searrow 0$
inevitably converges to a multiple of the identity. Consequently
such functions are not suitable to test the properties of states in
the scaling limit since they are not sensitive to their detailed
properties; they have the same limit in {\em every} state on
$\Al$.
For this reason we did not assume from the outset that the
elements of
$\Alu$
are continuous at
$\lambda = 0$.
As a consequence, the nets
$(\olu)_{\lambda > 0}$
are not convergent. To illustrate this fact consider for example
the operators
$\Cu$
of the form
$\Cu_{\lambda} = c_{\lambda}\cdot  1$, $c_{\lambda} \in \CC$,
                                             $\la > 0$.
There holds
$\olu(\Cu) = c_{\lambda}$,
hence if not all of the functions
$\lambda \to c_{\lambda}$
are continuous for
$\lambda \searrow 0$
the net
$(\olu)_{\lambda > 0}$
does not converge. Thus at this point we pay for the convenience
of not having to exhibit any particular renormalization group
transformation by admitting as elements of
$\Alu$
all functions which comply with the basic constraints deriving
from the renormalization group.

This apparent difficulty can be handled, however, with the help
of the Banach-Alaoglu theorem \cite{ReSi},
according to which every bounded set in the dual space of a Banach
space is pre-compact in the weak-*-topology. Applying this theorem
to the family of states
$(\olu)_{\lambda > 0}$
on the Banach space
$\Alu$
we see that this family contains (many) subnets which converge in
the
weak-*-topology for
$\lambda \searrow 0$.
This leads us to introduce the following terminology.
\\[6pt]
{\em Definition}: Let
$\omega$
be a state on
$\Al$
and
$\ou$
its canonical lift on
$\Alu$.
Each weak-*-limit point of the net
$(\olu)_{\lambda >0}$
for
$\lambda \searrow 0$
is called a {\em scaling limit state} of
$\ou$.
The scaling limit states of
$\ou$
are denoted by
$\oziu$, $\iota \in \II$
where $\II$
   is some index set, and the set of all these states is denoted
by
$SL(\ou)$.
\\[6pt]
For the physical interpretation of the scaling limit states
$\oziu$
on
$\Alu$
we apply the same rules as in the case of states
         at finite scales: we proceed to the local nets
\begin{equation}
\Oo \to \Aoi(\Oo) \doteq \Alu^{\pziu}(\Oo) =
  \Alu(\Oo)/{\rm ker}(\pziu)\, ,
\end{equation}
where
$\pziu$
is the GNS-representation induced by
$\oziu$.
According to Proposition 3.4 these nets are to be regarded
as ``limits'' of the underlying nets
$\Al_{\lambda}$
at scale
$\lambda > 0$,
hence our simplified notation. Similarly, the states
\begin{equation}
\ooi \doteq {\rm proj}\,\oziu
\end{equation}
on
$\Aoi$
are limits of the states
$\omega_{\lambda} = \omega$
on the nets
$\Al_{\lambda}$, $\lambda > 0$.
Thus by proceeding from the underlying net to the scaling algebra we
have been able to define nets and states in the scaling limit
$\lambda = 0$.
The apparent ambiguities in this construction, which are reflected
in the appearance of the index set $\II$, will be discussed below.

We are primarily interested here in the structure of the nets
$\Aoi$
and states
$\ooi$
induced by physical states.
\\[6pt]
{\em Definition:\/} A state
$\omega$
on the underlying net
$\Al$
is said to be a {\em physical state} if it is locally normal with
respect to the underlying vacuum representation (cf.\ relation (2.8)).

\vspace{6pt}
An important ingredient in our short distance analysis is the
following lemma, due to Roberts \cite{Ro1},
which is based on a result by
Wightman \cite{Wigh}.
              For the convenience of the reader we sketch the
proof of this statement.
\begin{Lemma} Let
$\omega_1$, $\omega_2$
be physical states on the net
$\Al$.
Then for any bounded region
$\Oo$ one has
$$
  ||\, (\omega_{1} - \omega_{2})\,|\,\Al(\lambda \Oo)\,||
  \to 0 \quad {\rm for} \quad \lambda \to 0\, , $$
where the given expression denotes the norm distance of the states,
restricted to the respective algebra.
\end{Lemma}
{\em Proof:\/} The proof proceeds in two steps. One first shows that
the intersection of the weak closures of all local algebras
centered at the origin $0$ of Minkowski space, say, is trivial,
$$ \bigcap_{\Oo \owns 0} \AO^{-} = \CC \cdot 1\,. $$
This is accomplished as follows: if $Z$ is any element in this
intersection the same is true for $Z^*$ and, since both operators are
localized at $0$, there holds
$[Z^{*},\ax(Z)] = 0$
for (strictly) spacelike $x$ because of locality. But
$x \to \ax (Z)$
is continuous in the weak operator topology, hence the
commutator vanishes also for lightlike $x$ and $x = 0$. We pick any
lightlike vector $e$ and consider the function
$\RR \owns t \to (\Omega,Z^{*}\alpha_{te}(Z)\Omega)$,
where
$\Omega$
is the vacuum vector in the defining representation of
$\Al$.
Its Fourier transform has support in
$\RR^{+} \cup \{0\}$
because of the relativistic spectrum condition. Because of
``lightlike commutativity'', the function coincides with
$t \to (\Omega,\alpha_{te}(Z)Z^{*}\Omega)$
whose Fourier transform has support in
$\RR^{-} \cup \{0\}$.
Since the intersection of these two sets is
$\{0\}$
and the function is bounded, it has to be constant. It follows that
$U(te)Z\Omega = Z\Omega$
for any
$t \in \RR$
and all lightlike vectors $e$. Consequently
$U(x)Z\Omega = Z\Omega$
for all
$x \in \RR^4$
and therefore
$Z\Omega = (\Omega,Z\Omega)\cdot \Omega$,
by the uniqueness of the vacuum vector. Since
$\Omega$
is separating%
\footnote{A vector $\Omega$ is called separating for a set
${\frak B}$ of bounded operators if, for each
$B \in {\frak B}$, $B\Omega =0$ implies $B = 0$.} for
the local algebras $\AO^{-}$ one arrives at
$Z=(\Omega,Z\Omega)\cdot 1$,
as claimed.

To complete the proof of the lemma suppose that
$\lkk$, $k \in \NN$,
is a sequence tending to $0$ such that for some
$\delta > 0$,
$||\,(\omega_{1} - \omega_{2})\,|\,\Al(\lkk\Oo)\,|| \geq \delta$.
Then there exist operators
$A_{k} \in \Al(\lkk\Oo)$
of norm $1$ such that
$|\,\omega_{1}(A_{k})-\omega_{2}(A_{k})\,| \geq \delta/2$, $k \in \NN$.
Since the set of operators
$A_k$, $k \in \NN$,
is bounded, it is pre-compact in the weak operator topology.
Its limit (accumulation) points Z are contained in
$\bigcap_{\Oo \owns 0}\AO^- $
since for any given open region
$\Oo_{0}$
containing $0$ there holds
$\Al(\lkk\Oo) \subset \Al(\Oo_{0})^-$
for sufficiently small
$\lkk$,
and
$\Al(\Oo_{0})^-$
is closed in the weak operator topology. Hence
$Z \in \CC \cdot  1$.
On the other hand, since $Z$ is a weak limit point of the set of
operators
$A_k$, $k \in \NN$,
and the states
$\omega_{i}$, $i =1,2$,
are locally normal with respect to the vacuum representation,
there exists an index
$l$
such that
$|\,\omega_{i}(A_{l}) - \omega_{i}(Z)\,| \leq \delta/8$, $i=1,2$.
But this implies that
$|\,\omega_{1}(Z) - \omega_{2}(Z) \,| \geq \delta/4$
which is a contradiction in view of the fact that $Z$ is a multiple
of the identity and  states are normalized. $\Box$
\\[6pt]
The following result is an immediate consequence of this lemma.
It says that all physical states look alike in the scaling limit.
\begin{Corollary} Let
$\omega_1$, $\omega_2$
be physical states on
$\Al$
and let
$\ou_{1}$, $\ou_{2}$
be their canonical lifts on
$\Alu$.
Then
$$ \lim_{\lambda \searrow 0}
 \, ||\,(\ou_{1,\lambda} - \ou_{2,\lambda})\,|\,\Alu(\Oo)
     \,|| = 0
      $$
for all bounded spacetime regions
$\Oo$.
In particular,
$SL(\ou_{1}) = SL(\ou_{2})$.
\end{Corollary}
{\em Proof:\/} Let
$\Oo$
be any bounded region and let
$\Au \in \Alu(\Oo)$.
By definition
$$ (\ou_{1,\lambda} - \ou_{2,\lambda})(\Au)
   = \omega_{1}(\Aul) - \omega_{2}(\Aul)\,, \quad \lambda > 0\,.
     $$
But
$\Aul \in \Al(\lambda\Oo)$,
so the first part of the statement follows from the
preceding lemma. Since the local algebras
$\Alu(\Oo)$, $\Oo \subset \RR^4$,
are norm dense in
$\Alu$,
this result implies that
$\lim_{\lambda \searrow 0}\,(\ou_{1,\lambda}
 - \ou_{2,\lambda})(\Au) = 0 $ for all
$\Au \in \Al$.
Hence the sets
$SL(\ou_{1})$
and
$SL(\ou_{2})$
coincide.   $\Box$
\\[6pt]
In view of the preceding arguments the following remark may be
in order: although the limit points of the nets
$\Aul$, $\lambda > 0$,
are multiples of the identity, this does  not imply that the
theory becomes trivial in the scaling limit. For these sequences
converge in general only in the weak operator topology. So the
limits of products
$\Aul \cdot \Bu_{\lambda}$
need not coincide with the products of the respective limits of
the factors
$\Aul$
and
$\Bu_{\lambda}$.
In this way non-trivial correlations between observables can
(and do) persist in the scaling limit.

According to the previous result it suffices for the
investigation of the scaling limit theories to concentrate
on the analysis of a single physical state. In the subsequent
lemma we make use of this fact and exhibit the structure of the
scaling limit states inherited from the underlying vacuum state.
\begin{Lemma} Let
$\omega$
be a physical state on
$\Al$
and let
$\ou$
be its canonical lift on
$\Alu$.
Then each scaling limit state
$\oziu \in SL(\ou)$
is a pure, Poincar\'e invariant vacuum state on
$\Alu$.
\end{Lemma}
{\em Remark:\/} This statement holds in any number of spacetime
dimensions greater than two. In two dimensional spacetime the
scaling limit states are also vacuum states, but they need not be
pure in general.
\\[6pt]
{\em Proof:\/} Without restriction of generality we may assume that
$\omega$
is the underlying vacuum state. In order to prove that the
corresponding states
$\oziu \in SL(\ou)$
are pure vacuum states on
$\Alu$
it suffices to show that they are invariant under the action of
the Poincar\'e transformations
$\auP$
and that the respective correlation functions have the right
continuity, support (in momentum space) and clustering
properties. The proof that the corresponding GNS-representations
are vacuum representations is then similar to the one in
standard quantum field theory, cf.\ \cite{StrW},
                                          and may be omitted here.

Let
$\oziu \in SL(\ou)$
and let
$(\ou_{\lk})_{\kappa \in {\KKf}}$
be a subnet of
$(\ou_{\lambda})_{\lambda > 0}$
which converges in the weak-*-topology to
$\oziu$, i.e., $\lim_{\kappa}\,\ou_{\lk}(\Au) = \oziu(\Au)$
for all
$\Au \in \Alu$.
Since each
$\ou_{\lambda}$
is a vacuum state on
$\Alu$,
cf.\ relation (3.16) and Corollary 3.3, there holds
$$  \oziu(\aLxu(\Au)) = \lim_{\kappa}\,\ou_{\lk}(\aLxu
 (\Au)) = \lim_{\kappa}\,\ou_{\lk}(\Au) = \oziu(\Au)
    $$
for all
$\Au \in \Alu$
and
$\Lx \in \Pg$.
The required continuity properties of the functions
$\Lx \to \oziu(\Bu\,\aLxu(\Au))$
follow from the norm continuity of the operators
$\Au \in \Alu$
with respect to Poincar\'e transformations, cf.\ relation (3.10).
To verify that
$\oziu$
induces a representation of
$\Alu$
which fulfills the relativistic spectrum condition we have
to show that
\begin{equation}
\int dx\,f(x)\,\oziu(\Bu\,\axu(\Au)) = 0
\end{equation}
for all
$\Au,\Bu \in \Alu$
and all functions
$f \in L^{1}(\RR^{4})$  whose Fourier transforms
have support in
$\RR^{4} \backslash \overline{V}_{+} $.
In view of the  continuity of the action of
$\auP$ and the fact that $\underline{\alpha}_{f}(\Au) \in \Alu$
one gets
$$ \int dx\,f(x)\,\oziu(\Bu\,\axu(\Au)) =
  \oziu(\Bu\,\underline{\alpha}_{f}(\Au)) =
  \lim_{\kappa}\,\ou_{\lk}(\Bu\,\underline{\alpha}_{f}(\Au))\,.$$
Making again use of the fact that each
$\ou_{\lambda}$
is a vacuum state on
$\Alu$,
it follows that the latter expression is equal to $0$ if
${\rm supp}\,\fT \subset \RR^{4}\backslash \overline{V}_{+}$.
This proves relation (4.4).

It remains to be shown that
$\oziu$
is a pure state on
$\Alu$,
i.e., that it satisfies the clustering condition. This we do as
follows. Let
$\Oo_r$
be the double cone with base radius
$r>0$, centered at the origin, and let
$\Au,\Bu \in \Alu(\Oo_r)$
be such that the functions
$t \to \underline{\alpha}_{t}(\Au)$
   and
$t \to \underline{\alpha}_{t}(\Bu)$
(where
$\underline{\alpha}_{t}$, $t \in \RR$,
denotes the translations in time direction) are
differentiable in the norm topology. The set of all
such operators (for arbitrary
$r > 0$)
is norm dense in
$\Alu$.
Now since each
$\ou_{\lambda}$
is a {\em pure} vacuum state on
$\Alu$
it follows from \cite[Eq.\ 3.4]{AHR} that for spatial translations
${\bf x}$
with
$|{\bf x}| >3r$
there holds
\begin{eqnarray*}
\lefteqn{|\, \ou_{\lk}(\Au\,\underline{\alpha}_{{\bf x}}
(\Bu)) - \ou_{\lk}(\Au)\cdot\ou_{\lk}(\Bu)\,| } \\
& & \leq\ c\frac{r^3}{|{\bf x}|^{2}} \cdot
 \left( \ou_{\lk}(\Au\,\Au^{*})^{1/2}\,\ou_{\lk}
 (\dot{\Bu}^{*}\dot{\Bu})^{1/2} +
 \ou_{\lk}(\Bu\,\Bu^{*})^{1/2}
 \, \ou_{\lk}(\dot{\Au}^{*}\dot{\Au})^{1/2} \right) \\
& & \leq\ c\frac{r^3}{|{\bf x}|^{2}} \cdot
  \left( ||\,\Au\,||\,||\,\dot{\Bu}\,|| +
        ||\,\dot{\Au}\,||\,||\,\Bu\,|| \right)\,,
\end{eqnarray*}
where $c$ is a universal (model independent) numerical constant
and the dot denotes time derivative. (Note that the relevant
inequality in \cite{AHR} contains a misprint.) The upper bound in
this estimate vanishes as
$|{\bf x}|^{-2}$
for
$|{\bf x}| \to \infty$,
uniformly in $\kappa$. This shows that the limit state
$\oziu$
is clustering for all pairs of operators
$\Au,\Bu$
described above. Since these operators are norm dense in
$\Alu$
it follows that
$\oziu$
is clustering and hence pure. $\Box$
\\[6pt]
It is an immediate consequence of this lemma that the kernels
of the GNS-representations
$\pziu$ of $\Alu$
induced by the  states
$\oziu \in SL(\ou)$
are invariant under Poincar\'e transformations. Hence, as discussed
in Sec.\ 3, there is an automorphic action
$\aoi_{\Pg}$
of the Poincar\'e group on each local net
$\Aoi$,
given by equation (3.20). Moreover, it  follows from the lemma that
the projected states
$\ooi$
are pure vacuum states on
$\Aoi$,$\aoi_{\Pg}$.
Thus all physical states look like vacuum states in the scaling
limit.
\\[6pt]
{\em Definition:\/} The nets $\Aoi,\aoi_{\Pg}$
are called {\em scaling limit nets} and the
corresponding states $\ooi$ {\em scaling limit states},
$\iota \in \II$.
\\[6pt]
       We summarize the results of this discussion
    in the following proposition.
\begin{Proposition} For $\iota \in \II$, let
$\Aoi,\aoi_{\Pg}$ and $\ooi$
be the scaling limit nets and scaling limit
                              states derived  from some physical
state $\omega$ on the underlying net
$\Al$.
Each
$\Aoi$
is a local net which is covariant with respect to the Poincar\'e
transformations
$\aoi_{\Pg}$
and each
$\ooi$
is a pure vacuum state on
$\Aoi,\aoi_{\Pg}$.
The collection of scaling limit nets and states does not depend
on the choice of the underlying physical state $\omega$.
\end{Proposition}
Let us turn now to the discussion of the ambiguities involved in
the definition of the scaling limit of a theory.
As was explained, these
ambiguities have their origin in the fact that the nets of
scaled states
$(\ou_{\lambda})_{\lambda >0}$
do not converge since the scaling algebra
$\Alu$
subsumes the orbits of local operators under the action of an
abundance of renormalization group transformations. Loosely
speaking, every scaling limit net
$\Aoi,\aoi_{\Pg}$
and state
$\ooi$
may be attributed to some particular choice of such a
transformation. Since the choice of renormalization group
transformations should not affect the physical interpretation
of the theory at small scales one may however expect that all
scaling limit theories describe the {\em same physics} at small
scales in generic cases. Bearing in mind that different,
but physically equivalent nets can be identified with the help
of net isomorphisms (cf.\ Sec.\ 3) we are led to the following
definition.
\\[6pt]
{\em Definition:\/} The underlying theory is said to have a {\em unique
scaling limit} if all scaling limit nets
$(\Aoi,\aoi_{\Pg})$, $\iota \in \II$,
derived from physical states on
$\Al$
are isomorphic. If there exist  net isomorphisms which
connect also the respective vacuum states
$\ooi$,
the theory is said to have a {\em unique vacuum structure} in the
scaling limit.
\\[6pt]
Making use of this terminology we see that, in
view of Proposition 4.4, there are the following principal
possibilities for the structure of the scaling limit of a
theory.

\newpage
\noindent
{\em Classification:\/} Let
$(\Aoi,\aoi_{\Pg})$, $\iota \in \II$,
be the scaling limit nets arising from physical states in the
underlying theory. There are the following alternatives: \\
i) All (global)  algebras
     $\Aoi$, $\iota \in \II$,
     consist of multiples of the identity. The theory is then said to
     have a {\em classical scaling limit}.   \\
ii)  All  nets
     $(\Aoi,\aoi_{\Pg})$, $\iota \in \II$,
     are isomorphic and non-abelian. The theory then has a
     {\em unique quantum scaling limit}. \\
iii) Not all of the  nets
     $(\Aoi,\aoi_{\Pg})$, $\iota \in \II$,
     are isomorphic. The theory is then said to have a
     {\em degenerate scaling limit}.  \\
${}$\\
Let us comment on the physical significance of these various
possibilities. We refer to the first case as ``classical'' since it
corresponds to the situation where the observables
$\Aul$
at small scales
$\lambda$
attain sharp (i.e., non-fluctuating) values in all physical states,
as is the case for classical observables in pure states. It is
note-worthy that the ``classical ideal'' in
$\Alu$
(generated by all commutators) is a proper ideal in such theories.
We would like to point out, however, that this classical
behaviour of the scaling limit with respect to renormalization
group transformations may have its origin in an exceptional
quantum behaviour of the observables in small regions:
due to the uncertainty principle, the energy-momentum transfer
of observables in
$\Al(\lambda\Oo)$
has to be at least of order
$\lambda^{-1}$
at small scales. But it could be substantially larger and behave
for  all observables in
$\Al(\lambda\Oo)$
like
$\lambda^{-q}$,
say, for some
$q > 1$.
Then it would not be possible to exhibit sequences of operators
$\Aul$
which occupy a limited volume of phase space at small scales,
apart from multiples of the identity (cf.\ the remark after
Lemma 2.2). Hence the scaling limit would be ``classical'' in such
theories. Examples where Planck's constant in the
uncertainty principle is effectively ``running'' in this way at
small scales could be non-renormalizable theories.

The second case corresponds to theories which have (in the
language of the renormalization group) an ultraviolet fixed
point. The simplest example of this kind is free field theory.
The corresponding nets can be shown to have a unique scaling
limit which, as  expected, is the net generated by a massless
free field \cite{BV}. This example, as trivial as it may be,
illustrates the fact that the apparent ambiguities in our
construction of the scaling limit disappear if one identifies
isomorphic nets. Of particular interest in this second class
of models are the asymptotically free theories. In such theories
the underlying quantum fields have, according to the folklore,
almost canonical dimensions (modified by logarithmic
corrections) and become massless free fields in the scaling
limit. We therefore propose to characterize such theories in the
present algebraic setting by the condition that the ``Huygens
ideal'' in
$\Alu$
(generated by the commutators of all pairs of operators which
are localized in strictly timelike separated regions) is
annihilated in this limit. Another class of theories with a
unique quantum scaling limit will be considered in the
subsequent section.

The last possibility according to our classification appears
if the structure of the underlying theory at small scales
cannot be described in terms of a single theory since it
varies continuously if one approaches
$\lambda =0$.
One then arrives by our construction at a collection of
non-isomorphic scaling limit theories which describe these
various structures. It is of interest that even in this complicated
situation the physical states can be interpreted at small scales
as vacuum states. But the properties of these vacua depend
on the scale.

In the remainder of this section we focus attention on the
physically significant class of theories with a unique
(quantum) scaling limit. It turns out that, due to this uniqueness,
dilations are a geometrical symmetry of the scaling limit nets.
\begin{Proposition} For $\iota \in \II$, let
$\Aoi,\aoi_{\Pg}$
be the (isomorphic) scaling limit nets in a theory with
unique scaling limit. There exist automorphisms
$\doim$, $\mu \in \RR^+$,
of the respective nets such that
$$  \doim (\Aoi(\Oo)) = \Aoi(\mu\Oo) $$
for all bounded regions
$\Oo$
and
$\mu > 0$,
and
$$ \doim \ckl \aoi_{\Lambda,x} = \aoi_{\Lambda,\mu x} \ckl \doim $$
for all
$\mu > 0$
and
$\Lx \in \Pg$.
Moreover, if the underlying theory has a unique vacuum structure
in the scaling limit there holds
$\ooi \ckl \doim = \ooi$
for all
$\mu > 0$.
\end{Proposition}
{\em Proof:\/} Since for any physical state
$\omega$
in the underlying theory and any
$\mu >0$
the corresponding sets
$(\ou_{\lambda})_{\lambda > 0}$
and
$(\ou_{\mu \lambda})_{\lambda >0}$
of lifted and scaled states on
$\Alu$
coincide, they have the same limit points. Hence the set
$SL(\ou)$
of scaling limit states is invariant under the (adjoint) action
of the scaling transformations
$\underline{\sigma}_{\mu}$.
The statement then follows from the assumption that all states
in
$SL(\ou)$
induce isomorphic nets and that there exist net isomorphisms
connecting the scaling limit states if the theory has a unique
vacuum structure in the scaling  limit.

To see this, let
$\oziu \in SL(\ou)$ and consider
the corresponding net
$\Aoi,\aoi_{\Pg}$ which is (by construction)
canonically isomorphic to
$\pziu(\Alu),{\rm Ad}\,\Uziu(\Pg)$.
Here
$(\pziu,\Hziu,\Omiu)$
is the GNS-representation induced by
$\oziu$ and $\Uziu$ the unitary representation of $\Pg$
on
$\Hziu$
(cf.\ Lemma 4.3). Now
$\ou_{\oim} \doteq \oziu \ckl \underline{\sigma}_{\mu}
   \in SL(\ou) $
and the corresponding net
$\Al_{\oim},\alpha^{\oim}_{\Pg}$
is canonically isomorphic to the net
$\pu_{\oim}(\Alu),{\rm Ad}\,\underline{U}_{\oim}(\Pg)$
                                           on $\Hziu$,
where
$\pu_{\oim} = \pziu \ckl \underline{\sigma}_{\mu}$
and
$\underline{U}_{\oim}\Lx = \Uziu(\Lambda,\mu x)$
for
$\Lx \in \Pg$.
Moreover, the state
$\ou_{\oim}$
is represented in this setting by the vector
$\Omiu \in \Hziu$.
Since the nets induced by
$\oziu$
and
$\ou_{\oim}$
are isomorphic (by the uniqueness of the scaling limit) there
exists a net isomorphism
$\toim$
mapping the net
$\pziu(\Alu),{\rm Ad}\,\Uziu(\Pg)$
onto the net
$\pu_{\oim}(\Alu),{\rm Ad}\,\underline{U}_{\oim}(\Pg)$.
More explicitly, there holds
$$ \toim(\pziu(\Alu(\Oo))) = \pu_{\oim}(\Alu(\Oo)) =
   \pziu(\Alu(\mu\Oo)) $$
for all bounded regions
$\Oo$,
and
$$ \toim \ckl {\rm Ad}\,\Uziu\Lx = {\rm Ad}\,\underline{U}_{\oim}
                                                      \ckl \toim
= {\rm Ad}\, \Uziu(\Lambda,\mu x) \ckl \toim $$
for all
$\Lx \in \Pg$ and $\mu >0$.
Thus each
$\toim$
induces an automorphism of the net
$\pziu(\Alu),{\rm Ad}\,\Uziu(\Pg)$
which has the geometrical interpretation of dilations.
Moreover, if the theory has a unique vacuum structure in the
        scaling limit there holds
$(\Omiu,\toim(\,.\,)\Omiu) = (\Omiu,\,.\,\Omiu)$
for some such
$\toim$.
Pulling back
$\toim$
to the net
$\Aoi,\aoi_{\Pg}$
with the help of the canonical isomorphism mentioned above one
obtains automorphisms
$\doim$
with the desired properties.  $\Box$
\\[6pt]
{\em Remark:\/} The automorphisms
$\doim$, $\mu \in \RR^+$,
in this statement do not necessarily provide a representation
of the multiplicative group
$\RR^+$.
But, as is easily checked, there holds
$$  \delta^{(0,\iota)}_{\lambda} \ckl \doim
    = \delta^{(0,\iota)}_{\lambda \mu} \ckl
    \zeta^{(0,\iota)}(\lambda,\mu)
                                  \,,\quad \lambda,\mu \in \RR^{+}, $$
where
   $\zeta^{(0,\iota)}(\,.\,,\,.\,)$
is a 2-cocycle on
$\RR^+$
with values in the group of internal symmetries of the net
$\Aoi,\aoi_{\Pg}$.
It is thus a problem of cohomology whether one can find a true
representation of the dilations by automorphisms of the scaling
limit theory. However, we do not enter into this question here.
\section{Scale invariant theories}
\setcounter{equation}{0}
 In the conventional
field theoretic setting, dilation invariant theories
are invariant (fixed points) under renormalization
group transformations.
In this section we show that these theories are also
invariant under the corresponding transformations
in the present setting. This result displays the
consistency of our method with the standard approach.

Let us begin by listing the additional assumptions
on the underlying theory made in this section.
\\[6pt]
{\em Dilation invariance:\/} On the underlying net
$\Al,\alpha_{\Pg}$
acts a family of automorphisms
$\delta_{\mu}$, $\mu \in \RR^+$,
such that
\begin{equation}
\delta_{\mu}(\AO) = \Al(\mu  \Oo)\,, \quad \mu>0
\end{equation}
for all bounded regions
$\Oo$
and
\begin{equation}
\delta_{\mu} \ckl \aLx = \alpha_{\Lambda,\mu x} \ckl
\delta_{\mu}\,, \quad \mu > 0
\end{equation}
for all
$\Lx \in \Pg$.
Moreover, the underlying vacuum state
$\omega(\,.\,) = (\Omega,\,.\,\Omega)$
on
$\Al$
is dilation invariant,
\begin{equation}
\omega \ckl \delta_{\mu} = \omega\,, \quad \mu > 0\,.
\end{equation}
Note that these assumptions coincide with the results
which we have established in Proposition 4.5 for theories
with a unique vacuum structure in the scaling limit. In
our subsequent analysis of the scaling limit of
dilation invariant theories we make in addition the following
assumption.
\\[6pt]
{\em Haag-Swieca compactness condition:\/}
Let
$E(\,.\,)$
denote the projections of the spectral resolution of
$U\,|\,\RR^4$,
let $\Delta$ be any compact subset of $\RR^4$ and let
$\Oo$
be any bounded region. Then the linear map from
$\AO$
into
$\cal H$
given by
\begin{equation}
A \to E(\Delta)A\Omega\,, \quad A \in \AO
\end{equation}
is a compact map between the Banach spaces $\AO$
       and $\Hh$.  (For a review of
compactness and nuclearity conditions, which impose certain
physically motivated constraints on the phase space properties
of a theory, cf.\ \cite{BuPo} and references quoted there.)
\\[6pt]
Let us now discuss the implications of the existence of the
dilations
$\delta_{\RRf^{ +}}$
in the underlying theory. Given any vector state in the underlying
vacuum representation, we consider its scaled canonical lift
on
$\Alu$
and the resulting net
$\Alu^{\pul}$, $\alpha^{\pul}_{\Pg}$.
As has been shown in Proposition 3.4, this net is isomorphic
to the underlying net
$\Al_{\lambda}$, $\alpha^{(\lambda)}_{\Pg}$
at scale $\lambda$. Now the existence of dilations implies that
the latter nets are isomorphic for different values of
$\lambda$, say
$\lambda_1$ and $\lambda_2$,
a net isomorphism being given by
$\delta_{\lambda_2 \lambda_1^{-1}}$,
cf.\ relations (5.1) and (5.2). Hence the nets
$\Alu^{\pul}$, $\alpha^{\pul}_{\Pg}$,
obtained from the original state by renormalization group
transformations, describe the same physics at each scale,
i.e., the theory is invariant (a ``fixed point'') under the
action of the renormalization group.

                                     It is not quite as simple to
show that this invariance under the renormalization group
persists also in the scaling limit. As a matter of fact,
we have been able to establish this result only under the
additional mild phase-space constraints imposed by the
Haag-Swieca compactness condition. We remark that the existence
of dilation invariant theories satisfying this condition has
been established in \cite{BuJa}.
\begin{Proposition}
Suppose that the underlying net
$\Al$, $\alpha_{\Pg}$
is dilation invariant and satisfies
the Haag-Swieca compactness condition. Then this
net has a unique scaling limit and vacuum structure in this limit.
In fact, the scaling limit nets
$\Aoi$, $\aoi_{\Pg}$, $\iota \in \II$,
are all isomorphic to the underlying net $\Al,\alpha_{\Pg}$
and the scaling limit states $\ooi$ are mapped onto the underlying
vacuum state $\omega$
by the respective net isomorphisms.
\end{Proposition}
{\em Proof:\/}
Let
$\ou$
be the canonical lift on
$\Alu$
of the underlying vacuum state $\omega$
on
$\Al$,
let
$\oziu \in  SL(\ou)$
and let
$\Aoi$, $\aoi_{\Pg}$
be the corresponding scaling limit net.
As in the proof of Proposition 4.5 it is convenient
to make use of the fact that this net is canonically
isomorphic to
$\pziu(\Alu)$, ${\rm Ad}\,\Uziu(\Pg)$,
where
$(\pziu,\Hziu,\Omiu)$
is the GNS-representation of
$\Alu$
induced by
$\oziu$.

Let
$(\ou_{\lambda_{\kappa}})_{\kappa \in \KKf}$
be a subnet of
$(\ou_{\lambda})_{\lambda > 0}$
which converges in the weak-*-topology to
$\oziu$.
A net isomorphism $\phi$
mapping $\pziu(\Alu), {\rm Ad}\, \Uziu(\Pg)$ onto
$\Al,\alpha_{\Pg}$
  is obtained by setting for
$\Au \in \bigcup_{\Oo}\Alu(\Oo)$
\begin{equation}
\phi(\pziu(\Au)) \doteq w-\lim_{\kappa}\,
\delta_{\lambda_{\kappa}}^{-1}(\Au_{\lambda_{\kappa}})\,.
\end{equation}
We thus have to prove the following statements:
(a) The right hand side of (5.5)
exists and defines (b) a linear map from the algebra
$\pziu(\Alu)$
into
$\cal B(H)$
which (c) is *-preserving. (d) The map $\phi$ intertwines
the action of the Poincar\'e transformations
${\rm Ad}\,\Uziu(\Pg)$
and
$\alpha_{\Pg}$
and there holds for all bounded regions
$\Oo$, $\phi(\pu(\Alu(\Oo))) = \AO$.
Finally (e), $\phi$
is multiplicative.
\\[4pt]
(a) Let
$\Oo$
be any bounded region and let
$\Au \in \Alu(\Oo)$.
To simplify notation we set
$\Ak \doteq \delta_{\lambda_{\kappa}}^{-1}(\Au_{\lambda_{\kappa}})$
and note that
$\Ak \in \AO$
and
$||\, \Ak \,|| \leq ||\, \Au\,||$
for
$\kappa \in \KK$.
Let
$B \in \Al(\Oo_1)$,
where
$\Oo_1$
is any other bounded region.
We first show that the net of numbers
$(\omega(B\,\Ak))_{\kappa \in \KKf}$
converges: it follows from relations (5.1) and (5.2)
that the function
$\lambda \to \Bu_{\lambda} \doteq \delta_{\lambda}(B)$
is an element of
$\Alu(\Oo_1)$.
Hence, using the invariance of $\omega$ under dilations we obtain
$$
\omega(B\,\Ak) = \omega \ckl \delta_{\lambda_{\kappa}}
(B\,\Ak) = \omega(\Bu_{\lambda_{\kappa}}\,\Au_{\lambda_{\kappa}})
= \ou_{\lambda_{\kappa}}(\Bu\,\Au) $$
and consequently the limit exists,
$$
\lim_{\kappa}\, \omega(B\,\Ak) = \oziu(\Bu\,\Au)\,. $$
Since the vacuum vector $\Omega$ is cyclic for the algebra
$\bigcup_{\Oo_1}\Al(\Oo_1)$
and the family of operators
$\Ak$, $\kappa \in \KK$,
is uniformly bounded we conclude that
$w-\lim_{\kappa}\,\Ak \Omega$
exists. As $\Omega$ is separating for the weak closures
$\AO^-$
of the local algebras it follows that also
$w-\lim_{\kappa}\, \Ak$
exists and that the limit is an element of
$\AO^-$.
\\
(b) For the proof that the map $\phi$ is well defined
and linear it suffices to show that for
$\Au \in \Alu(\Oo)$, $\pziu(\Au) = 0$
implies
$w-\lim_{\kappa}\, \Ak \Omega = 0$
and consequently
$w-\lim_{\kappa} \Ak = 0$
(recall that $\Omega$ is separating for
$\AO^-$).
Now, making again use of the fact that $\omega$ is invariant
under dilations, there holds
\begin{eqnarray*}
 \lefteqn{||\,\pziu(\Au)\Omiu\,||^{2}  =
\lim_{\kappa}\,\ou_{\lambda_{\kappa}}(\Au^{*}\,\Au)} \\
 & &=\ \lim_{\kappa}\,\omega(\Au_{\lk}^{*}\,\Au_{\lk}^{*})
 = \lim_{\kappa}\,\omega(\delta_{\lk}^{-1}(\Au_{\lk}^{*})
                       \delta_{\lk}^{-1}(\Au_{\lk})) \\
& & =\ \lim_{\kappa}\, ||\, \Ak\Omega\,||^2\, ,
\end{eqnarray*}
which establishes the desired implication.
\\
(c) That $\phi$ is *-preserving, i.e.,
$$
    \phi(\pziu(\Au)^{*}) = \phi(\pziu(\Au))^{*}\, ,
    $$
follows from the fact that
$w-\lim_{\kappa}\,\Ak = A$
implies
$w-\lim_{\kappa}\,\Ak^{*} = A^* $.
\\
(d) The proof that $\phi$ intertwines the respective
actions of the Poincar\'e transformations is accomplished
by noting that for any
$\Lx \in \Pg$
and any
$\Au \in \bigcup_{\Oo} \Alu(\Oo)$
there holds
\begin{eqnarray*}
\lefteqn{\phi(\Uziu\Lx\pziu(\Au)\Uziu\Lx^{-1}) =
   \phi(\pziu(\aLxu(\Au))) } \\
& & =\ w-\lim_{\kappa}\,\delta_{\lk}^{-1}((\aLxu(\Au))_{\lk})
= w-\lim_{\kappa}\,\delta_{\lk}^{-1} \ckl \alpha_{\Lambda,
\lk x}(\Au_{\lk}) \\
& & =\ w-\lim_{\kappa}\,\alpha_{\Lambda,x}
 \ckl \delta_{\lk}^{-1}(\Au_{\lk})
= w-\lim_{\kappa}\, U\Lx \Ak U\Lx^{-1} \\
& & =\ U\Lx
  \left (    w-\lim_{\kappa}\,\Ak \right )
                                 U\Lx^{-1}
= \alpha_{\Lambda,x}\ckl \phi(\pziu(\Au))\,.
\end{eqnarray*}
{}From this chain of equalities it follows in particular that
$$
  ||\, \alpha_{\Lambda,x} \ckl \phi(\pziu(\Au)) -
        \phi(\pziu(\Au)) \,|| \leq ||\,\aLxu(\Au) -\Au \,||\, ,
   $$
hence all elements of
$\phi(\pziu(\Alu(\Oo)))$
are norm continuous with respect to the action of
$\alpha_{\Pg}$.
Since we already know from step (a) that
$\phi(\pziu(\Alu(\Oo))) \subset \AO^-$
we conclude that
$\phi(\pziu(\Alu(\Oo))) \subset \AO$
as
$\AO$
is, according to our assumptions in Section 2, the maximal
subalgebra of operators in
$\AO^-$
which are norm continuous under Poincar\'e transformations.
The opposite inclusion is obtained by noting, as in step (a),
that for
$B \in \AO$
the corresponding function
$\lambda \to \Bu_{\lambda} \doteq \delta_{\lambda}(B)$
is an element of
$\Alu(\Oo)$
and there holds
$\phi(\pu(\Bu)) = \lim_{\kappa}\,\delta_{\lk}^{-1}(\Bu_{\lk}) = B$.
Hence $\phi(\pziu(\Alu(\Oo))) = \AO$.
\\
(e) It remains to establish the multiplicativity of the map
$\phi$, and it is here where the Haag-Swieca compactness
condition enters. The crucial step in the argument is the
demonstration that the weakly convergent net of operators
$\Ak \in \AO$, $\kappa \in \KK$,
considered in step (a), is even convergent in the strong
operator topology. Again it suffices to establish the strong
convergence of
$\Ak \Omega$
since $\Omega$
is separating for
$\AO^-$
and the family
$\Ak$, $\kappa \in \KK$,
is uniformly bounded. We begin by noting that
\begin{eqnarray*}
\lefteqn{ ||\, \ax(\Ak) - \Ak \,|| = ||\, \ax\delta_{\lk}^{-1}
(\Au_{\lk}) - \delta_{\lk}^{-1}(\Au_{\lk}) \,|| }\\
 & & =\ ||\,\delta_{\lk}^{-1}(\alpha_{\lk x}(\Au_{\lk})
                                                    - \Au_{\lk})\,||
 \leq ||\, \axu(\Au) - \Au \,||
\end{eqnarray*}
showing that the functions
$x \to \ax(\Ak)$
are continuous at $x=0$, uniformly in
$\kappa \in \KK$.
Hence, as in the proof of  Lemma 3.1 one can
find for given $\varepsilon >0$ a compact set
$\OT \subset \RR^4$
and operators
$\widetilde{A}_{\kappa} \in \AOT$
such that
$||\, \Ak - \widetilde{A}_{\kappa} \,|| < \varepsilon$,
$\kappa \in \KK$.
Consequently
$$
 ||\, \Ak\Omega - E(\OT)\Ak\Omega \,||
 = ||\,( 1 - E(\OT))(\Ak - \widetilde{A}_{\kappa})\Omega \,||
< \varepsilon \, . $$
Hence in order to prove strong convergence of the net
$(\Ak\Omega)_{\kappa \in \KKf}$
it suffices to prove strong convergence of
$(E(\OT)\Ak\Omega)_{\kappa \in \KKf}$
for all compact regions
$\OT$.
Now for any such
$\OT$
the latter net forms, by the compactness condition, a strongly
precompact subnet of
$\cal H$
since
$\Ak \in \AO$
and
$||\,\Ak\,|| \leq  ||\, \Au\,||$, $\kappa \in \KK$.
Whence the weakly convergent nets
$(E(\OT)\Ak\Omega)_{\kappa \in \KKf}$
converge also strongly and consequently
$(\Ak\Omega)_{\kappa \in \KKf}$
converges strongly, as claimed. Now let
$\Au,\Bu \in \Alu(\Oo)$, $\Oo$
bounded, and let
$\Ak,\Bk \in \AO$
be the corresponding nets introduced in (a).
Since both nets are uniformly bounded and convergent in the strong
operator topology there holds
$$
(s-\lim_{\kappa}\,\Ak)\cdot(s-\lim_{\kappa}\,\Bk)\ =\
 s-\lim_{\kappa}\,\Ak\,\Bk \ = \
                             s-\lim_{\kappa}\, \delta_{\lk}^{-1}
 ((\Au \cdot \Bu)_{\lk})\, . $$
This shows that
 $$
\phi(\pziu(\Au)) \cdot \phi(\pziu(\Bu)) = \phi(\pziu(\Au \Bu))
= \phi(\pziu(\Au)\cdot \pziu(\Bu)) \, ,$$
i.e., $\phi$ is multiplicative. By continuity, $\phi$ can be extended
to a map from
$\pziu(\Alu)$
to
$\Al$, and it thus has all properties required for a net isomorphism.

The remaining statement about the action of $\phi$ on the
vacuum state $\omega$ follows from the following relation for
all
$\Au \in \Alu$,
$$
   \omega \ckl \phi(\pziu(\Au)) = \lim_{\kappa}\omega \ckl
   \delta_{\lk}^{-1}(\Au_{\lk}) = \oziu(\Au)\,, $$
where we made once again use of the fact that $\omega$
is invariant under dilations. This completes the proof of
the proposition. $\Box$
\\[6pt]
We emphasize that the latter result does {\em not} mean
that the nets
$(\ou_{\lambda})_{\lambda >0}$
of states on the scaling algebra converge in the scaling limit.
There still exists an abundance of different limit points
$\oziu$
on
$\Alu$.
But the result illustrates our statement that (a) all limit points
describe the same physics in generic cases and (b) that the choice
of renormalization group transformations does not affect the physical
interpretation.
\section{Geometric modular action and scaling limit}
\setcounter{equation}{0}
\noindent
The scaling limit nets and states provide information
about the properties of physical states at very small
scales. Of particular interest are the particle structure
and the symmetries appearing in this limit. In this section
we present a structural result which provides a basis for
such an analysis. We will show that if the underlying theory
satisfies a condition of geometrical modular action
invented by Bisognano and Wichmann, then the scaling limit
has this property, too. As a consequence, the scaling limit
nets comply with the condition of (essential) Haag
duality and one can apply the methods of Doplicher and Roberts
\cite{DR} to determine from these nets the charged fields and the
global gauge group appearing in the scaling limit.
These applications will be discussed elsewhere.

                                                Another
interesting consequence of this result is the insight that,
excluding the case of theories with a classical scaling
limit, the local von Neumann
                 algebras $\AO^-$ of the underlying nets
are of type ${\rm III}_1$ for any double cone $\Oo$.
This reproduces a theorem by Fredenhagen \cite{Fr} in the present
general setting.

We begin by explaining some notation. Introducing proper
coordinates on $\RR^4$, we choose two lightlike vectors
$e_{\pm} \doteq (\pm1,1,0,0)$ and consider the two
opposite wedge-shaped regions ${\cal W}_{\pm}$
(wedges, for short), given by
\begin{equation}
 \Wp = - \Wm \doteq \{x \in \RR^{4} : x \cdot e_{\pm}
      > 0 \} \,.
\end{equation}
These regions are mutually spacelike, $(\Wp)' = \Wm$,
and invariant under the one parametric group
$\Lambda_t$, $t \in \RR$, of proper Lorentz transformations
(``boosts'') fixed by $\Lambda_t e_{\pm} =
{\rm e}^{\pm t}e_{\pm}$. We also consider the discrete
Lorentz transformation $ j$ which satisfies
$ j e_{\pm} = - e_{\pm}$ and acts like the
identity on the edge $(e_{\pm})^{\perp}$ of the wedges
${\cal W}_{\pm}$.    Thus $ j^2 =  1$
and $ j\Wp = \Wm$.

Within the underlying theory we define the $C^*$-algebras
$\Al({\cal W}_{\pm})$ as inductive limits of all local
algebras $\AO$, where $\Oo$ is bounded and
$\overline{\Oo} \subset {\cal W}_{\pm}$. The definition of
the corresponding scaling algebras
$\Alu({\cal W}_{\pm}) \subset \Alu$ and of the algebras
$\Aoi({\cal W}_{\pm}) \subset \Aoi$ in the scaling limit
theory is analogous. The following auxiliary lemma, which
we record here for later use, summarizes some elementary
properties of the scaling limit states $\oziu$.
Throughout this section we assume that these states are
obtained from physical states in the underlying theory.
\begin{Lemma}
Let $\oziu$ be any scaling limit state on $\Alu$
and let
$(\pziu,\Hziu,\Omiu)$ be the corresponding GNS-representation.
\\
i) The vector $\Omiu$ is cyclic and separating for $\pziu(\Alu
       ({\cal W}_{\pm}))^-$.   \\
ii) If the representation is non-trivial, i.e., if the
dimension of $\Hziu$ is greater than one, then
$\pziu(\Alu({\cal W}_{\pm}))^-$ is a factor of
type ${\rm III}_{1}$.
\end{Lemma}
{\em Proof:\/} {\em i)\/} It follows from the relativistic
spectrum condition (cf. Lemma 4.3) and a well-known argument due to Reeh
and Schlieder \cite{RSchl} that
$$ \overline{
\pziu(\Alu({\cal W}_{\pm}))\Omiu}=    \overline{
\bigcup_{x}\pziu(\underline{\alpha}_{x}(\Alu({\cal W}_{\pm})))
             \Omiu}\,.
  $$
The set on the right hand side  contains
the vector states in $\pziu(\Alu(\Oo))\Omiu$ for all
bounded regions $\Oo$. Since $\Omiu$ is cyclic for
$\pziu(\Alu)$ by the GNS-construction, the cyclicity of
$\Omiu$ for $\pziu(\Alu({\cal W}_{\pm}))^-$ follows.
That $\Omiu$ is also separating for these algebras is then
a consequence of locality.  \\
{\em ii)\/} This statement is also based on the relativistic
spectrum condition which implies that the centralizer
of the algebras $\pziu(\Alu({\cal W}_{\pm}))^-$ relative to the
vacuum state consists of multiples of the identity. Cf.\
references \cite{Dr,Lo} for a proof. $\Box$
\\[6pt]
In the following we assume that the modular operators
(introduced in the Tomita-Takesaki theory \cite{BrRo}) which
are affiliated with the vacuum state $\Omega$ and the
algebra $\Al(\Wp)^-$ of the underlying theory act like
Lorentz transformations on the underlying net.
\\[6pt]
{\em Condition of geometric modular action:\/} Let
$\Delta, J$ be the modular operator and modular conjugation
associated with the pair $ \Al(\Wp)^-,\Omega $ in the
underlying theory. For each bounded region
$\Oo \subset \RR^4$ there holds
\begin{equation}
J\,\AO^-J = \Al( j\Oo)^-\,.
\end{equation}
Moreover, the modular objects $\Delta,J$ are related to the
underlying Poincar\'e transformations $U(\Pg)$ according to
\begin{equation}
     J\,U\Lx J = U( j\Lambda   j, jx)\,,
     \quad \Lx \in \Pg,
\end{equation}
and, with $\Lambda_t$ as defined above,
\begin{equation}
\Delta^{it} = U(\Lambda_{2\pi t})\,,\quad  t \in \RR\,.
\end{equation}

\vspace{6pt}
This tight relation between modular objects and geometrical
symmetries was first established by Bisognano and Wichmann
for local nets which arise from an underlying Wightman theory
     \cite{BiWi}.
In two dimensions the condition has also been established
in a purely algebraic setting \cite{Bo4}, and in physical spacetime
it is a consequence of much weaker assumptions \cite{BrGuLo, GuLo}.
Therefore, the condition may be viewed as being ``generic''
in quantum field theory. It is linked to the existence of a
PCT-operator $\Theta$ which is related to the modular
conjugation $J$ in the above condition
according to
\begin{equation}
J = U(R)\Theta\,,
\end{equation}
where $R$ denotes the rotation about the 1-axis by the
angle $\pi$.

If the underlying theory has this property there holds,
for any $A \in \AO$, $JAJ \in \Al( j\Oo)$ (since $\AO$ is the
maximal subalgebra in $\AO^-$ of operators which are
norm continuous under Poincar\'e transformations)  and, for any
$A \in \Al$, $J\aLx(A)J = \alpha_{ j\Lambda j,
 jx}(JAJ)$. Using these facts we introduce an
anti-automorphism $\xu : \Alu \to \Alu$ by setting
\begin{equation}
(\xu(\Au))_{\lambda} \doteq J\,\Aul J \,, \quad \lambda > 0\,.
\end{equation}
Relations (6.2) and (6.3) imply that
\begin{equation}
\xu(\Alu(\Oo)) = \Alu( j\Oo)
\end{equation}
 for all bounded regions $\Oo$  and
\begin{equation}
  \xu \ckl \aLxu = \underline{\alpha}_{ j\Lambda
   j, jx} \ckl \xu\,, \quad \Lx \in \Pg\,.
\end{equation}
It is also clear from its definition that $\xu$ commutes
with the scaling transformations,
\begin{equation}
\xu \ckl \smu = \smu \ckl \xu\,, \quad \mu > 0\,.
\end{equation}
Since $J$ is antiunitary and $J\Omega = \Omega$ there holds
$(\Omega,JAJ\Omega) = (\Omega,A^*\Omega)$, $A \in \Al$.
Hence, if $\omega(\,.\,) = (\Omega,\,.\,\Omega)$ denotes the
vacuum state and $\ou$ its canonical lift on $\Alu$ it
follows that
\begin{equation}
\ou \ckl \xu (\Au) = \ou(\Au^*) = \overline{\ou(\Au)}
\,, \quad  \Au \in \Alu\,,
\end{equation}
i.e., $\ou$ is skew-invariant under the adjoint action of
$\xu$.

Now let $\oziu \in SL(\ou)$ be any scaling limit
state on $\Alu$, let $(\pziu,\Hziu,\Omiu)$ be its
GNS-representation and let $\Uziu$ be the unitary
representation of $\Pg$ on $\Hziu$, cf.\ Lemma 4.3.
It is our aim to show that the corresponding scaling limit net
complies with the principle of geometric modular action.
To this end we consider the modular operator and conjugation
$\Dziu,\Jziu$ associated with the pair
$\pziu(\Alu(\Wp))^-,\Omiu$. (Note that these operators are
well-defined in view of the first part of Lemma 6.1.)
In the following crucial lemma we show that they are related to the
Poincar\'e transformations.
%
%
\begin{Lemma}
Assume that the underlying theory satisfies the condition
of geometric modular action. Then it holds that
$$ \Dziu^{it} = \Uziu(\Lambda_{2\pi t})\,, \quad t \in \RR\,,
          $$
where $\Lambda_t$, $t \in \RR$, are the proper Lorentz transformations
introduced above, and
$$ \Jziu \Uziu\Lx \Jziu = \Uziu( j\Lambda j,
       jx)\,, \quad \Lx \in \Pg\,.
      $$
Moreover, the modular conjugation $\Jziu$ implements the
action of $\xu$ in the representation $\pziu$,
$$ {\rm Ad}\,\Jziu \ckl \pziu = \pziu \ckl \xu\,. $$
\end{Lemma}
{\em Proof:\/} We use again the fact that $\oziu$ is
the weak-*-limit of some convergent subnet $(\ou_{\lk})
_{\kappa \in \KKf}$ of the lifted and scaled vacuum state
$\omega$. Let $\Au,\Bu \in \Alu(\Wp)$ and consider the
functions
$$ t \to \ou_{\lk}(\Au\,\underline{\alpha}_{\Lambda_{t}}
(\Bu)) = (\Omega,\Au_{\lk}\alpha_{\Lambda_{t}}(\Bu_{\lk})\Omega) $$
for  $\kappa \in \KK$. Since $\Aul, \Bu_{\lambda}
\in \Al(\Wp)$ for all $\lambda >0$ it follows from relation
(6.4) and the fundamental results of the Tomita-Takesaki
theory \cite{BrRo} that the restriction of each state $\ou_{\lk}$
to the $C^*$-dynamical system $\Alu(\Wp),
\underline{\alpha}_{\Lambda_{\RRf}}$ satisfies the
KMS-condition at fixed inverse temperature $2\pi$.
As a consequence, the weak-*-limit point $\oziu$
of these functionals has this property, too
               \cite[Thm.\ 5.3.30]{BrRo}.
Hence the unitary group $\Uziu(\Lambda_{2\pi t})$,
$t \in \RR$, which implements the automorphisms $\underline{
\alpha}_{\Lambda_{2\pi t}}$ in the representation $\pziu$
is the modular group corresponding to
$\pziu(\Alu(\Wp))^-,\Omiu$. The first equality in the lemma
then follows.

In the next step we prove the third equality. Because of the
relations (6.9) and (6.10) there holds
$\ou_{\lambda} \ckl \xu =
            \overline{\ou_{\lambda}}$, $\lambda > 0$,
and consequently the limit state $\oziu$ is also skew-invariant
under the action of $\xu$. Thus there exists on $\Hziu$ an
antiunitary operator $\Iu$ which implements the action
of $\xu$, i.e., $ \Iu\,\pziu(\Au) \Iu = \pziu(\xu(\Au))$
for all $\Au \in \Alu$,  and which leaves $\Omiu$ invariant.
We will show that $ \Iu = \Jziu$. To this end we pick
any $\Au \in \Alu$ and any $\Bu^* \in \Alu(\Wp)$ which
is entire analytic with respect to the action of
$\underline{\alpha}_{\Lambda_{\RRf}}$. Since $\Wp$ is
invariant under the action of the Lorentz transformations
$\Lambda_{\RRf}$ it is apparent that the set of all
such operators is norm  dense in $\Alu(\Wp)$. Now, as
$\underline{\alpha}_{\Lambda_{i\pi}}(\Bu^*) \in \Alu(\Wp)$,
we have
\begin{eqnarray*}
\lefteqn{
 (\Omiu,\pziu(\Au)\Uziu(\Lambda_{i\pi})\pziu(\Bu^*)\Omiu)
= (\Omiu,\pziu(\Au)\pziu(\underline{\alpha}_{\Lambda_{i\pi}}
    (\Bu^*))\Omiu)}\\
& &=\ (\Omiu,\pziu(\Au\,\underline{\alpha}_{\Lambda_{i\pi}}
    (\Bu^*))\Omiu)
    = \lim_{\kappa}\,(\Omega,\Au_{\lk}(\underline{\alpha}
    _{\Lambda_{i\pi}}(\Bu^*))_{\lk}\Omega) \\
& & =\ \lim_{\kappa}\,(\Omega,\Au_{\lk}\alpha_{\Lambda_{i\pi}}
    (\Bu_{\lk}^{*})\Omega)
 = \lim_{\kappa}\, (\Omega,\Au_{\lk}U(\Lambda_{i\pi})
     \Bu_{\lk}^*\Omega )  \\
& &=\  \lim_{\kappa}\,(\Omega,\Au_{\lk}J\Bu_{\lk}\Omega)
 = \lim_{\kappa}\,(\Omega,\Au_{\lk}(\xu(\Bu))_{\lk}\Omega) \\
& & =\  (\Omiu,\pziu(\Au\,\xu(\Bu))\Omiu)
       = (\Omiu,\pziu(\Au)\pziu(\xu(\Bu))\Omiu) \,,
\end{eqnarray*}
where in the sixth equality we made use of relation (6.4)
and the Tomita-Takesaki theory. Since $\Au \in \Alu$ was
arbitrary, it follows from the above equation and the
preceding result that
$$
  \Jziu\pziu(\Bu)\Omiu = \Uziu(\Lambda_{i\pi})\pziu(\Bu^*)
  \Omiu = \pziu(\xu(\Bu))\Omiu =  \Iu \, \pziu(\Bu)\Omiu \,.
  $$
But the set of vectors $\pziu(\Bu)\Omiu$, where $\Bu \in
\Alu(\Wp)$ is analytic, is dense in $\Hziu$, cf.\ Lemma 6.1,
                                             and since
$\Jziu$ and $\Iu$ are bounded operators it follows that
$ \Iu = \Jziu$, as claimed. The second equality in the
lemma is now an immediate consequence of relation (6.8). $\Box$
\\[6pt]
It follows from this result and the Tomita-Takesaki theory
that the scaling limit nets comply with the condition
of geometric modular action. For the precise formulation
of this result we have to consider the representations of
the nets $\Aoi,\aoi_{\Pg}$ which are induced by the corresponding
scaling limit
states $\ooi$, $\iota \in \II$. Since these representations
are the defining representations of the respective nets
we do not indicate in the following statement the pertinent
homomorphisms for the sake of a simpler notation.
\begin{Proposition}
Given an underlying theory which complies with the condition
of geometric modular action, let $\Aoi,\aoi_{\Pg}$ be the
associated scaling limit nets in the defining representation
induced by $\ooi$, and let $\Uoi(\Pg)$ and $\Omoi$ be the
respective Poincar\'e transformations and the GNS-vector
corresponding to $\aoi_{\Pg}$ and $\ooi$, $\iota \in \II$.
\\
(i) The modular operators $\Delta_{0,\iota}$
     and conjugations $J_{0,\iota}$
fixed by the pairs $\Al_{0,\iota}(\Wp)^-$, $\Omoi$ satisfy
relations (6.2) to (6.4) with reference to the respective
nets $\Aoi$ and representations $\Uoi$ of $\Pg$. \\
(ii) The nets $\Aoi$ satisfy wedge-duality, i.e.\
$$ \Aoi(\Wp)' = \Aoi(\Wm)^-  \,. $$
\end{Proposition}
{\em Proof:\/} Any net $\Aoi,\aoi_{\Pg}$ is,
by its very definition,
isomorphic to the net $\pziu(\Alu)$, ${\rm Ad}\,\Uziu(\Pg)
                                                     $ on $\Hziu$,
and the corresponding isomorphism connects the states
$\ooi$ and $(\Omiu,\,.\,\Omiu)$. The first part of the
statement then follows from the preceding lemma and the
geometric action of $\xu$ given in relation (6.7).
As a consequence,
$$ \Aoi(\Wm)^- = J_{0,\iota} \Aoi(\Wp)^-
                       J_{0,\iota} = \Aoi(\Wp)'\,,
   $$
where the second equality holds according to the Tomita-Takesaki
theory. This proves the second part of the statement.
$\Box$
\\[6pt]
This result shows that the property of geometric modular
action of the underlying theory persists in the scaling limit.
In view of Proposition 4.5 and results by Guido and Longo \cite{GuLo}
one may expect that this property can also directly be established
in the scaling limit in theories of physical interest, such as
asymptotically free theories. Its consequence, wedge duality and,
as a further implication, (essential) Haag duality \cite{Ro2}, is an
important ingredient in the Doplicher-Haag-Roberts approach
to the superselection analysis \cite{DHR}.
               As another interesting consequence
of wedge duality in the scaling limit we establish the following
result on the type of local algebras in the  underlying
theory.
\begin{Proposition}
Suppose that the underlying net $\Al$ has a non-classical
scaling limit, cf.\ Sec.\ 4, which satisfies wedge-duality.
Then the local von Neumann algebras $\AO^-$ are of type
${\rm III}_{1}$ for all double cones $\Oo$.
\end{Proposition}
{\em Proof:\/} We quote the following criterion for the
type ${\rm III}_{1}$-property, originally established
by Fredenhagen \cite{Fr}, from \cite[Sec.\ 16.2]{BauWo}.
\\[6pt]
{\em Criterion:\/} Let $\cal M$ be a von Neumann algebra on
some Hilbert space with separating vector\footnote{ In
\cite{BauWo} it was assumed that $\Omega$ is also cyclic for
$\cal M$. But this is not necessary since the criterion,
as stated here, implies that the stronger version is
satisfied in the subspace $\overline{{\cal M}\Omega}$.}
$\Omega$. The algebra $\cal M$ is of type ${\rm III}_{1}$
if, given $\vth \in \RR_{+}$ and $\varepsilon > 0$,
there exist nets $\Ak \in \cal M$, $\Bk \in {\cal M}'$,
$\kappa \in \KK$, such that
$$ ||\,\Ak\Omega\,|| \geq 1\,, \quad
   ||\,(\vth^{1/2}\Bk - \Ak^*)\Omega \,|| < \varepsilon\,,
   \quad ||\,(\Bk^* - \vth^{1/2}\Ak)\Omega \,||
   < \varepsilon
   $$
and $\Ak^*\Ak$, $\Bk^*\Bk$ and $\Bk\Ak$ tend weakly to
multiples of  1.
\\[6pt]
We shall exhibit operators $\Ak,\Bk$ with the required
properties by making use of the assumption that there
exists some scaling limit state $\oziu$ which induces
a non-trivial representation $(\pziu,\Hziu,\Omiu)$ of
$\Alu$. By the second part of Lemma 6.1 the algebras
$\pziu({\cal W}_{\pm})^-$ are then factors of type
${\rm III}_{1}$, and taking into account the condition
of wedge duality we get, using the Tomita-Takesaki theory,
$$ \Jziu\pziu(\Alu(\Wp))^-\Jziu = \pziu(\Alu(\Wp))' =
\pziu(\Alu(\Wm))^- \,,$$
where $\Dziu,\Jziu$ are the modular operator and conjugation
corresponding to the pair $\pziu(\Alu(\Wp))^-,\Omiu$.
We note that, by Kaplansky's density theorem \cite{KaRi},
         the domain
$\pziu(\Alu(\Wp))\Omiu$ is a core for the modular operator
$\Dziu^{1/2}$, and by taking into account that
$\Alu(\Wp)$ is generated by its local subalgebras, the same
holds true for the domain $\bigcup_{\overline{\Oo} \subset
\Wp} \pziu(\Alu(\Oo))\Omiu$, where the regions $\overline{\Oo}$ are
compact.

In view of these facts it follows from the very definition
of von Neumann algebras of type ${\rm III}_{1}$ that
for each $\vth \in \RR_+$ and $\varepsilon > 0$ there
exists an operator $\Au \in \Alu(\Oo_1)$, where
$\overline{\Oo_1} \subset \Wp$ is compact, such that
$$ ||\, (\pziu(\Au)^* - \vth^{1/2}\Jziu\pziu(\Au)\Jziu)
     \Omiu\, || = ||\,(\Dziu^{1/2} - \vth^{1/2})
     \pziu(\Au)\Omiu \,|| < \varepsilon
     $$
while
$$ || \pziu(\Au)\Omiu \,|| > 1 \,.$$
By wedge duality one has $\Jziu \pziu(\Au)\Jziu \in
\pziu(\Alu(\Wm))^-$, and applying Kaplansky's theorem a second
time we see that there exists an operator $\Bu \in
\Alu(\Oo_2)$, where $\overline{\Oo}_2 \subset \Wm$ is compact,
such that
$$  ||\, (\pziu(\Bu) - \Jziu\pziu(\Au)\Jziu)\Omiu\,|| <
\varepsilon \  {\rm and} \
    ||\, (\pziu(\Bu)^* - \Jziu\pziu(\Au)^*\Jziu)\Omiu \,||
    < \varepsilon . $$
Combining these estimates with the preceding inequalities
and making use of the fact that $\Jziu$ is an anti-unitary
involution, we arrive at
\begin{eqnarray*}
\lefteqn{
 ||\, (\pziu(\Au)^* - \vth^{1/2}\pziu(\Bu))\Omiu\,||
\leq (1+\vth^{1/2})\varepsilon
\quad {\rm and}} \\
 & & ||\, (\pziu(\Bu)^* - \vth^{1/2}\pziu(\Au))\Omiu\,||
\leq (1+\vth^{1/2})\varepsilon \,.
\end{eqnarray*}
Bearing in mind that the GNS-representation $(\pziu,\Hziu,
\Omiu)$ is induced by a scaling limit state $\oziu$
which in turn is the weak-*-limit point
of a net $(\ou_{\lk})_{\kappa \in \KKf}$ arising
from the underlying vacuum state, we conclude  that
there is some $\kappa_0' \in \KK$ such that for
$\kappa > \kappa_0'$ there holds
$||\,\Au_{\lk}\Omega\,|| \geq 1$ and
$$ ||(\Au_{\lk}^*- \vth^{1/2}\Bu_{\lk})\Omega\,||
\leq (1+\vth^{1/2})\varepsilon \,,
\quad
 ||(\Bu_{\lk}^*- \vth^{1/2}\Au_{\lk})\Omega\,||
\leq (1+\vth^{1/2})\varepsilon  \,.
   $$
Since $\Au \in \Alu(\Oo_1)$, $\Bu \in \Alu(\Oo_2)$,
we have also $\Au_{\lk} \in \Al(\lk\Oo_1)$
and $\Bu_{\lk} \in \Al(\lk\Oo_2)$.

Now let $\Oo$ be any double cone such that
$\Oo \subset \Wp$ and such that  the origin $0$ lies
on its boundary. Since $\overline{\Oo_1}$ is a compact
subset of the (open) wedge $\Wp$, it follows that
$\lk \Oo_1 \subset \Oo$ for sufficiently small
$\lk$, i.e., for all $\kappa > \kappa_0''$
with some suitable $\kappa_0'' \in \KK$. Hence
$\Au_{\lk} \in \AO^-$ for all of these $\kappa$. On the
other hand, there holds $\lk \overline{\Oo_2} \subset
\Wm$ for all $\kappa$ and consequently
$\Bu_{\lk} \in \Al(\Wm)^- \subset \AO'$, where the
inclusion follows from locality. Finally, since
$\overline{\Oo_1}$ and $\overline{\Oo_2}$ are compact,
the weak limits of the nets
$\Au_{\lk}^*\Au_{\lk}$, $\Bu_{\lk}^*\Bu_{\lk}$ and
$\Au_{\lk}\Bu_{\lk}$ are elements of the algebra
$\bigcap_{\Oo \owns 0} \AO^-$ and therefore multiples
of the identity (cf.\ the proof of Lemma 4.1). Thus,
picking $\kappa_0 > \kappa_0',\kappa_0''$ we see
that the nets
$\Ak \doteq \Au_{\lk} \in \AO^-$ and
$\Bk \doteq \Bu_{\lk} \in \AO'$, $\kappa > \kappa_0$,
have the properties required by the criterion.
Hence $\AO^-$ is of type ${\rm III}_{1}$. But every
double cone can be brought by Poincar\'e transformations
into the particular position required in the preceding
argument, proving the statement of the proposition. $\Box$
\section{Outlook}
\setcounter{equation}{0}
\noindent
In the present article we have established a general
method for the short distance analysis of local nets
of observables. Based on the idea that the
choice of renormalization group transformations
identifying physical observables at different scales
ought to be largely arbitrary, apart from a few basic
constraints regulating their phase space properties,
we were led to the concept of scaling algebra. These
scaling algebras are canonically associated with any
given net and combine in a convenient way the information
about the theory at different scales. It turned out
that the scaling algebras have again the structure of local,
Poincar\'e covariant nets on which the renormalization
group acts by automorphisms (scaling transformations).
These features greatly simplify the short distance
analysis since they allow it to apply  well-known
methods and results from algebraic quantum field theory. Several
arguments given in Sections 4 and 6 illustrate this
point.

In a forthcoming paper it will be shown how, by using these
methods, the phase space properties of the underlying theory
                                    which can be expressed
in terms of nuclearity and compactness conditions (cf.\
 \cite{BuPo}
and references quoted there), determine
the phase space structure in the scaling limit.
This point is of relevance on one hand for the question
of the general nature of the scaling limit (``classical''
versus ``quantum''). On the other hand it is an important
ingredient for its physical interpretation since the
phase space properties of a net reveal the causal and
thermal features of a theory as well as its particle
aspects, cf.\ \cite{Ha}.

The next step in our programme is the systematic analysis
of the super\-selection and particle structure of the
scaling limit theories. Roughly speaking, these theories
play a similar role as the asymptotic free field theories
in scattering theory: they provide information about the
stable particle content and the symmetries of the theory
which become visible in the respective limit; yet they do
not contain by themselves any information about the
interacting dynamics. This information is gained by
{\em comparing} the properties of the theory at finite
scales with those in the scaling limit.

                                        The ultraparticles
of a theory, i.e., the particle structures appearing
at very small scales such as partons, leptons etc., can
be identified with the particles (in the sense of Wigner)
of the scaling limit theory \cite{BuBS}. Of course, information
about the masses of the ultraparticles gets lost in the
scaling limit. But the comparison of the ultraparticle
content with the particle content of the underlying
theory allows one to decide whether a particle is confined
at finite scales, i.e., exists only as an ultraparticle,
or disappears in the scaling limit since it disintegrates,
say, into non-interacting ultraparticles, as in the case
of hadrons.  Thus physical ideas
and concepts which so far have been studied only in specific
models become accessible in our
setting to a more general analysis.

Of similar interest is the related problem of the superselection
structure of the scaling limit theories. Anticipating that
the underlying theory complies with the condition of geometric
modular action, given in Sec.\ 6, we have shown that the
scaling limit satisfies the condition of essential
Haag duality. Hence, by the fundamental work of Doplicher
and Roberts \cite{DR} we know that the superselection structure
of any scaling limit theory is in one-to-one correspondence
to the spectrum (that is, the dual) of a compact group
$G_0$. Moreover, there exist charged Bose- and Fermi fields which
transform as tensors under the action of $G_0$ and generate
the charged scaling limit states from the scaling limit
vacuum. Again, the symmetry group $G$ and the field content
of the underlying theory may be quite different from that of its
scaling limit. In asymptotically free theories, cf.\ Sec.\ 4,
one would expect that the underlying symmetry group
$G$ becomes larger in the scaling limit since interactions
which can conceal a symmetry at finite scales (confinement)
are turned off in the scaling limit. Yet also the opposite
phenomenon may occur, e.g., in theories with a classical
scaling limit.

The structure of the gauge group $G_0$ in the scaling limit
is of particular interest in asymptotically free gauge
theories since it should contain information about the
type of the underlying
        {\em local} gauge group. Since in the scaling limit
                                       there is no difference
between global and local gauge transformations,
and all charge degrees of freedom should become visible
in case of asymptotic freedom, the local gauge group
``at a point'' should appear as a subgroup of $G_0$.
Thus the scaling algebra is a promising tool for uncovering,
from the gauge invariant observables, the full gauge group
of a theory.

For the sake of concreteness let us illustrate the above
points by the example of the Schwinger model, i.e.,
massless quantum electrodynamics
in two spacetime dimensions, see \cite{BLTO,Stroc}
                                             and references quoted
therein. As is well-known, the algebra of observables of this
theory is generated (in the defining vacuum representation)
                    by the free massive scalar field $\phi$,
and all finitely localized observables are elements of the
local von Neumann algebras generated by $\phi$, or are
affiliated with them in the case of unbounded operators.
There exists a conserved current in this model,
$j_{\mu} = \partial^{\nu}\varepsilon_{\mu \nu}\phi$,
but there does not exist any charged superselection sector
of states with finite energy,  cf. for example
\cite[Thm. 3.1(1)]{FMS}. In fact, if one forgets
about the origin of the local net of observables in this
model, there does not seem to be any particular reason to
relate it to a gauge theory.

The picture changes, however, if one proceeds to small
scales with the help of the scaling algebra. It turns out
that the scaling limit of the theory is a local extension
of the net generated by the free scalar massless field
(in exponentiated Weyl form). Let us mention as an aside that
the use of bounded operators pays off at this point since
one does not run into the familiar infrared problems
connected with the massless scalar field in two dimensions.
Now the interesting point is that the scaling limit theory
has a one-parameter family of superselection sectors
in the sense of Doplicher, Haag and Roberts which carry a
non-trivial charge with respect to the scaling limit of the
current $j_{\mu}$. Thus one encounters in the scaling limit
``electrically'' charged states which do not appear at
finite scales, as is expected in asymptotically free
theories exhibiting confinement. It is also worth noting
that the vacuum state in the scaling limit is not pure
anymore, but the algebra of observables attains a center.
This degeneracy of the vacuum is akin to the so-called
$\theta$-vacuum structure of the model at finite scales.
Thus we recover by our general method from the net of local
observables the well-known features of the Schwinger model
which are believed to illustrate the structure of physically
more interesting theories, such as quantum chromodynamics
\cite{BeJo}. A detailed account of these results, as well as a
discussion of other simple models, where the scaling limit nets
can be computed explicitly, will be published elsewhere.

After the clarification of the possible structure of the
scaling limit and its comparison with the underlying theory
it would be desirable to understand the more detailed
short distance properties of physical states. As we have
seen in the present article, for $\lambda \searrow 0$ the
leading contribution of the scaled states $\ou_{\lambda}$
on the scaling algebra $\Alu$ are vacuum states. In next to
leading order there ought to appear the ultraparticle content
of the given states \cite{BuBS}. Thus by an asymptotic expansion
of the functionals $\ou_{\lambda}$ with respect to $\lambda$
one should be able to establish a general
characterization of those physical states which may be
viewed as bound states of ultraparticles.

We conclude this outlook with the remark that the scaling
algebra provides also a setting for the discussion of the infrared
limit $\lambda \to \infty$. Moreover, the general ideas underlying
the construction of the scaling algebra should also be
useful for the investigation of other scaling limits,
such as the non-relativistic limit $c \to \infty$ or the
classical limit $\hbar \to 0$. Thus the concept of
scaling algebra seems to be a promising tool to tackle a
number of interesting physical problems within the setting
of algebraic quantum field theory.
\newpage
\noindent
{\Large {\bf Appendix}}
\\[18pt]
\renewcommand{\theequation}{\Alph{section}.\arabic{equation}}
\setcounter{section}{1}
\setcounter{equation}{0}
\noindent
The ideas of the renormalization group have
proven to be useful in the general analysis of quantum
field theories on curved spacetimes as well. They are, for
example, an important ingredient in the formulation
of the principle of  local stability, proposed
in \cite{HNS,FrHa} to chararacterize physical states in those cases
where the underlying spacetime structure does not
exhibit any symmetries admitting the definition of
vacuum states. The renormalization group has also been
used to derive the possible values of the Hawking
temperature from first principles \cite{HNS,Hes} and to determine
the type of local von Neumann algebras \cite{BauWo,Wol}. All these
interesting results rely on the existence of quantum
fields. It is the aim of the present Appendix to
indicate how the concept of scaling algebra
may serve as a substitute in investigations of the
short distance properties of states and observables
on arbitrary spacetime manifolds $\cal M$ with
Lorentzian metric $g$.

Let $({\cal M},g)$ be a (four-dimensional) spacetime;
to simplify the discussion we assume that $({\cal M},g)$
is globally hyperbolic (cf.\ \cite{ONei} for precise
definitions). The open, relatively compact subsets
(regions) of $\cal M$ will be denoted by $\Rr$.
Two regions $\Rr_1,\Rr_2$ are said to be {\em spacelike
separated} if there does not exist any causal curve
with respect to the metric $g$ which connects $\Rr_1$
and $\Rr_2$.

A local quantum theory on $({\cal M},g)$ is fixed by
specifying a net $\Rr \to {\cal A}(\Rr)$ of unital
$C^*$-algebras on $\cal M$ which satisfies the
conditions of isotony and locality (spacelike commutativity).
We assume that this net is concretely given on some
Hilbert-space of ``physical states''.

It is our aim to analyze the observables and physical states
of the underlying theory at  small scales.
Within the present setting changes of the spatio-temporal
scale can conveniently be described in tangent space:
let $q \in \cal M$ be any spacetime point which will be
kept fixed in the following and let $T_{q}\cal M$
be the tangent space of $\cal M$ at $q$. The open subsets
with compact closure of $T_{q}\cal M$ will be
denoted by $\Oo$. We fix a (sufficiently small, starshaped)
neighbourhood $\Oo_q$ of $0 \in T_q\cal M$ and map it
by the exponential map ${\rm exp}_q$ onto a neighbourhood
$\Rr_q$ of $q \in \cal M$.
Similarly, we identify the subsets $\Oo \subset \Oo_q$
with the regions $\Rr = {\rm exp}_q(\Oo) \subset \Rr_q$.
This identification of regions in $\cal M$ and $T_q\cal M$
by the use of the exponential map is suggested by the idea
that physics at very small scales should be describable
in terms of Minkowski space, in accordance with the
equivalence principle of general relativity. As is well known,
the exponential map complies with this idea in the best
possible way: its inverse ${\rm exp}_{q}^{-1}$ induces, from
the underlying metric $g$, a metric on (a suitable neighbourhood
of 0 in) $T_q\cal M$ which coincides with the usual
Minkowskian metric at the origin $0 \in T_q\cal M$, and
deviates from it in a neighbourhood of $0$ only up to
terms of {\em second} order, in geodesic normal coordinates
at $q$. We also recall that ${\rm exp}_{q}$ maps the
straight lines through $0$ to the geodesics passing
through $q$.

These geometrical facts suggest that for the purpose of
describing the underlying theory at small scales in the
neighbourhood $\Rr_q$ of $q$ it is natural to proceed
from the net $\cal A$ on $\cal M$ to the associated
net $\Al$ on the tangent space $T_q\cal M$ given by
\begin{equation}
 \Oo \to \AO \doteq {\cal A}({\rm exp}_{q}
       (\Oo \cap \Oo_q))\,, \quad \Oo \subset T_q\cal M
\end{equation}
if $\Oo \cap \Oo_q \neq \emptyset$. If
$\Oo \cap \Oo_q = \emptyset$, we put $\AO = \CC\cdot 1$.
It is easily checked that the assignment (A.1)
is isotonous in $\Oo$ and hence it defines indeed a net
on $T_q\cal M$ with $C^*$-inductive limit $\Al$,
describing the possible observations in $\Rr_q$.
We emphasize that the net $\Al$ dependes on the choice
of the spacetime point $q \in \cal M$, the neighbourhood
$\Rr_q \subset \cal M$ (respectively, $\Oo_q \subset T_q\cal M$)
and, of course, the chosen, physically motivated identification
of the regions $\Oo \subset \Oo_q$ and $\Rr \subset \Rr_q$
via the exponential map. But we refrain from indicating
this dependence in the notation.

The geometrical aspects of  renormalization group transformations
can now be described as in the case of theories on Minkowski
space. We consider uniformly bounded
functions $\lambda \to \Aul$, $\lambda > 0$ with values in
the algebra of observables $\Al$ and equip them with the
algebraic structure and $C^*$-norm given in relations
(3.7) and (3.8). Given any subset $\Oo \subset T_q\cal M$,
we define the corresponding local algebra by
\begin{equation}
 \Alu(\Oo) \doteq \{ \Au : \Aul \in \Al(\lambda \Oo),
                    \,    \la > 0  \} \,.
\end{equation}
The assignment $\Oo \to \Alu(\Oo)$  defines a net
on $T_q\cal M$ with corresponding global algebra $\Alu$
on which scaling transformations $\smu$, $\mu \in \RR^+$,
can be defined as in relation (3.14). Hence, lifting the underlying
states $\omega$ on $\Al$ to the algebra $\Alu$ by setting
\begin{equation}
\ou(\Au) \doteq \omega(\Au_{\lambda = 1})\,,
\end{equation}
we can again describe in a convenient way the observations
made at small scales by considering the scaled states
$\ou \ckl \sglu$ for $\lambda \searrow 0$.

The algebra $\Alu$ contains all functions which comply with
the geometrical aspects of renormalization group transformations.
But since we did not yet impose any constraints on their
phase-space properties, not all of these functions can be
interpreted as orbits of local operators under renormalization
group transformations (keeping the unit of $\hbar$ fixed).
We will indicate below how one can get rid of these unwanted
elements by proceeding to suitable subnets of $\Alu$.

Before entering into that discussion let us mention that
already in the present very general setting one can
determine the geometrical and causal properties of the nets
appearing in the scaling limit: let $\omega$ be any state
on $\Al$, let $\ou$ be its canonical lift on $\Alu$ and
let $\oziu$ be any limit point of the net $(\ou \ckl \sglu)_
{\lambda >0}$ for $\lambda \searrow 0$. It turns out that
the corresponding net $\Oo \to \Aoi(\Oo) \doteq
\Alu(\Oo)/{\rm ker}(\pziu)$, where $\pziu$ is the
GNS-representation induced by $\oziu$, can be interpreted
as a local net in Minkowski space if one defines on
$T_q\cal M$ the Minkowski metric with respect to any
normal geodesic coordinate system at $q$. Moreover,
the scaling limit nets so obtained are independent of the
 choice of the neighbourhood $\Oo_q \subset T_q\cal M$
entering into the definition of $\Al$. This result
shows that our approach is fully consistent with the equivalence
principle.

Let us now turn to the problem of selecting elements of
$\Alu$ which can be regarded as orbits under the action
of the renormalization group. Again it may seem natural
to characterize the functions by the requirement that they
occupy at each scale $\lambda$ a fixed volume of
``phase space''. But the implementation of this idea is
not quite as simple as in  Minkowski space since there
need not exist any isometries of $({\cal M},g)$, i.e.,
energy-momentum need not be conserved. Thus, selecting
elements of $\Alu$ by the condition that they have at
each scale $\lambda > 0$ a specific energy-momentum
transfer involves a certain degree of arbitrariness,
as energy-momentum depends in general on the choice of
a Cauchy-surface.

If one is primarily interested in the properties of the
theory in the scaling limit one can circumvent this
problem, however. Namely, it suffices then to control the
energy-momentum transfer of the operators $\Aul$
in the limit $\lambda = 0$, where they are localized at
the spacetime point $q$.
This ``local'' point of view is also suggested by
classical field theory, where phase space appears
as cotangent bundle of the underlying configuration
space.

To implement this idea let us assume that the underlying
theory admits (locally) a ``dynamics''. This requires
that there exists a smooth foliation of the region
$\Rr_q$ by Cauchy-surfaces ${\cal C}(t)$, where
$-t_0 < t < t_0$ denotes the time-parameter (the time function
$\Rr_q \to \RR$ being appropriately normalized), and
$q \in {\cal C}(0)$. The dynamics is then given by a
family of endomorphisms $\alpha_{t_{1},t_{2}}$,
$-t_0 < t_2 \leq t_1 < t_0$, of the net $\cal A$.
They map the local algebras ${\cal A}(\Rr)$ into
${\cal A}(\Rr_q)$ if $\overline{\Rr} \subset \Rr_q$ and
$t_1,t_2$ are sufficiently small and satisfy the equation
\begin{eqnarray}
\alpha_{t_{1},t_{2}} \ckl \alpha_{t_{2},t_{3}}
          & = & \alpha_{t_{1},t_{3}} \\
        \alpha_{t,t} &=& {\rm id}\,. \nonumber
\end{eqnarray}
Let us briefly explain the physical significance of the
endomorphisms $\alpha_{t_{1},t_{2}}$ (sometimes called
propagators). In suitable representations of ${\cal A}$
one may think of $\alpha_{t_{1},t_{2}}$ as time ordered
exponential (Dyson expansion) of generators
$\delta_t(\,.\,) = i\,[H_t,\,.\,]$, where $H_t$ can
locally be interpreted as a Hamiltonian on the Cauchy-surface
${\cal C}(t)$. The operators $H_t$ in turn may be thought of
in generic cases as suitably regularized integrals of a
local stress energy tensor, integrated over ${\cal C}(t)$.
We need not enter here into the subtle mathematical
problems appearing in the rigorous construction of these
quantities and only mention that local dynamics have been
constructed in models for globally hyperbolic spacetimes,
cf.\ \cite{Kay}.

Let us now return to our problem of characterizing
elements of $\Alu$ with sufficiently regular phase
space properties at small scales. Picking any dynamics,
and recalling that the fixed point $q$ under scaling
transformations lies on the Cauchy-surface ${\cal C}(0)$,
it seems natural to test the energy momentum transfer of the
operators $\Aul$ at small scales with the help of the
endomorphisms $\alpha_{t,0}$ for small $t$. Thus we propose
to select the desired elements of $\Alu$ by the condition
\begin{equation}
\limsup_{\la \searrow 0}\,||\,\alpha_{\lambda t,0}(\Aul)
  - \Aul \,|| \to 0 \quad {\rm for} \quad t \to 0\,,
\end{equation}
in analogy to condition (3.5). Thinking of the generators
of the dynamics in terms of Cauchy-surface integrals of a
local stress energy tensor, it is evident that condition (A.5)
does not depend on the shape of ${\cal C}(0)$ in the spacelike
complement of $q$. So the arbitrariness in selecting regular
elements from $\Alu$ largely disappears. By imposing
condition (A.5) for the dynamics corresponding to all
possible choices of a time direction we arrive at a strictly
local generalization of condition (3.5).

It is apparent that condition (A.5) determines a subnet
of $\Alu$ which is larger than its analogue introduced in the
case of Minkowski space. For one does not impose any restrictions
on the energy momentum transfer of the underlying operators
at finite scales. But this is irrelevant for the discussion
of the scaling limit where the properties of the elements
$\Au \in \Alu$ at finite scales are not tested. To illustrate
this fact let us note that if one imposes in the case of
Minkowski space theories conditons (3.5) and (3.6) on the
elements of the scaling algebra only in the limit of small
$\lambda$, as in relation (A.5), one arrives at the same
scaling limit(s) of the underlying theory as in our present
treatment.

We conclude this Appendix with the remark that there exist
interesting variants of condition (A.5). For example, one may
select ``smooth'' elements of $\Alu$ by imposing the condition
\begin{equation}
\limsup_{\lambda \searrow 0} \,
 ||\, \lambda\cdot \delta_{t=0}(\Aul)\,|| < \infty \,,
\end{equation}
where $\delta_t$ are the generators of the dynamics (propagators).
Quite another approach to testing the phase space
properties of $\Alu$ can be based on the modular groups
associated with suitable physical states and local
algebras, cf.\ \cite[Sec.\ 5]{BuDALo}. A more detailed exposition
of the formalism of scaling algebras on arbitrary spacetime
manifolds as well as some applications will be presented
elsewhere.
\\[24pt]
{\Large {\bf Acknowledgement}}
\\[18pt]
\noindent
We gratefully acknowledge financial support by the DFG (Deutsche
Forschungsgemeinschaft).


\begin{thebibliography}{99}
%
\bibitem{Ha}
Haag, R.:
{\em Local Quantum Physics.
                                                                  }
Berlin, Heidelberg, New York: Springer 1992
%
%
\bibitem{Am}
Amit, D.J.:
{\em Field theory, the renormalization group, and critical
phenomena.                                                        }
Singapore: World Scientific 1984
%
%
\bibitem{Bo1}
Borchers,\ H.J.:
{\em \"Uber die Mannigfaltigkeit der interpolierenden Felder zu einer
kausalen S-Matrix.                                                }
Nuovo Cimento {\bf 15} (1960) 784
%
%
\bibitem{HaSw}
Haag, R., Swieca, J.A.:
{\em When does a quantum field theory describe particles?
                                                                  }
Commun.\ Math.\ Phys.\ {\bf 1} (1965) 308
%
%
\bibitem{BiWi}
Bisognano,\ J.J., Wichmann,\ E.H.:
{\em On the duality condition for a hermitean scalar field.
                                                                  }
J.\ Math.\ Phys.\ {\bf 16} (1975) 985   and
J.\ Math.\ Phys.\ {\bf 17} (1976) 303
%
%
\bibitem{Fr}
Fredenhagen,\ K.:
{\em On the modular structure of local algebras of observables.}
Commun.\ Math.\ Phys.\ {\bf 97} (1985) 79
%
%
\bibitem{SuWe}
Summers,\ S.J.,\ Werner,\ R.F.:
{\em  Maximal violation of Bell's inequalities for algebras
of observables in tangent spacelike regions.                     }
Ann. Inst. H. Poincar\'e {\bf 49} (1988) 215
%
%
\bibitem{BuDaFr}
Buchholz,\ D.,\ D'Antoni,\ C.,\ Fredenhagen,\ K.:
{\em The universal structure of local algebras.             }
Commun.\ Math.\ Phys.\ {\bf 111} (1987) 123
%
%
\bibitem{DHR}
Doplicher, S., Haag, R., Roberts, J.E.:
{\em Fields, observables and gauge transformations.}
Commun. Math. Phys. {\bf 13} (1969) 1 and
                  Commun. Math. Phys.  {\bf 15} (1971) 173
%
\bibitem{Bo2}
Borchers, H.J.:
{\em Local rings and the connection of spin with statistics.
                                                                  }
Commun.\ Math.\ Phys.\ {\bf 1} (1965) 291
%
%
\bibitem{Di}
Dixmier, J.: {\em  $C^*$-Algebras.} Amsterdam: North-Holland 1977
%
\bibitem{ReSi}
Reed, M., Simon, B.: {\em Methods of modern mathematical
physics, Vol.\ 1.} New York: Academic Press 1975
%
\bibitem{Ro1}
Roberts, J.E.: {\em Some applications of dilatation
invariance to structural questions in the theory
of local observables. } Commun.\ Math.\ Phys. {\bf
37} (1974) 273
%
\bibitem{Wigh}
Wightman, A.S.: {\em La th\'eorie quantique locale et la
th\'eorie quantique des champs.} Ann. Inst. H. Poincar\'e
{\bf 1} (1964) 403
%
\bibitem{StrW}
Streater, R.F., Wightman, A.S.: {\em PCT, spin and statistics,
and all that.} New York: Benjamin 1964
%
\bibitem{AHR}
Araki, H., Hepp, K., Ruelle, D.: {\em On the asymptotic
behaviour of Wightman functions in spacelike directions.}
Helv. Phys. Acta {\bf 35} (1962) 164
%
\bibitem{BV}
Buchholz, D., Verch, R.: In preparation
%
\bibitem{BuPo}
Buchholz, D., Porrmann, M.: {\em How small is the phase
space in quantum field theory? } Ann.\ Inst.\ H.\
Poincar\'e {\bf 52} (1990) 237
%
\bibitem{BuJa}
Buchholz, D., Jacobi, P.: {\em On the nuclearity condition
for massless fields.} Lett.\ Math.\ Phys.\ {\bf 13} (1987) 313
%
\bibitem{DR}
Doplicher, S., Roberts, J.E.: {\em Why there is a field algebra
with a compact gauge group describing the superselection
structure in particle physics.} Commun.\ Math.\ Phys.\
{\bf 131} (1990) 51
%
\bibitem{RSchl}
Reeh, H., Schlieder, S.: {\em Bemerkungen zur Unit\"ar\"aquivalenz
von Lorentzinvarianten Feldern.} Nouvo Cimento {\bf 22} (1961)
1051
%
\bibitem{Dr}
Driessler, W.: {\em Comments on lightlike translations
and applications in relativistic quantum field theory.}
Commun.\ Math.\ Phys.\ {\bf 44} (1975) 133
%
%
\bibitem{Lo}
Longo, R.: {\em Notes on algebraic invariants for
non-commutative dynamical systems.}
Commun.\ Math.\ Phys.\ {\bf 69} (1979) 195
%
\bibitem{BrRo}
Bratteli, O., Robinson, D.W.: {\em Operator algebras
and quantum statistical mechanics, Vol.\ 2.}
New York, Berlin, Heidelberg: Springer-Verlag 1986
%
\bibitem{Bo4}
Borchers, H.J.: {\em The CPT-theorem in two-dimensional
theories of local observables.} Commun.\ Math.\ Phys.\
{\bf 143} (1992) 315
%
\bibitem{BrGuLo}
Brunetti, R., Guido, D., Longo, R.:
{\em Modular structure and duality in conformal
quantum field theory.}
Commun. Math. Phys. {\bf 156} (1993) 201
%
\bibitem{GuLo}
Brunetti, R., Guido, D., Longo, R.:
{\em Group cohomology, modular theory and spacetime
symmetries.} To appear in Rev. Math. Phys.
%
%
\bibitem{Ro2}
Roberts, J.E.: {\em Lectures on algebraic quantum field
theory.} in:  {\em The algebraic theory of
superselection sectors.} Kastler, D., ed.
                         Singapore: World Scientific 1990
%
\bibitem{BauWo}
Baumg\"artel, H., Wollenberg, M.:
{\em Causal nets of operator algebras. } Berlin: Akademie Verlag
1992
%
%
%
\bibitem{KaRi}
Kadison, R.V., Ringrose, J.R.: {\em Fundamentals of the theory
of operator algebras, Vol. 2.} Orlando: Academic Press 1986
%
\bibitem{BuBS}
Buchholz, D.: {\em On the manifestations of particles,} in:
 {\em Mathematical physics
towards the 21st century.} Sen, R.N., Gersten, A., eds.
                           Beer-Sheva: Ben-Gurion University Press
1994
%
\bibitem{BLTO}
Bogolubov, N.N., Logunov, A.A., Oksak, A.I., Todorov, I.T.:
{\em General principles of quantum field theory.}
Dordrecht: Kluwer Academic Publishers 1990
%
\bibitem{Stroc}
Strocchi, F.: {\em Selected topics on the general properties
of quantum field theory.} Singapore: World Scientific 1993
%
\bibitem{FMS}
Fr\"ohlich, J., Morchio, G., Strocchi, F.:
{\em Charged sectors and scattering states in quantum
electrodynamics.} Ann. Phys. (N.Y.) {\bf 119} (1979) 241
%
\bibitem{BeJo}
Becher, P., Joos, H.: {\em 1+1 dimensional quantum
electrodynamics as an illustration of the hypothetical
structure of quark field theory.} in:
 {\em Proceedings of the 5th meeting on
fundamental physics.} De Guevara, P.L. et al., eds.
                      Madrid: J.E.N. 1977
%
\bibitem{HNS}
Haag, R., Narnhofer, H., Stein, U.: {\em On quantum field
theory in gravitational background.} Commun.\ Math.\ Phys.\
{\bf 94} (1984) 219
%
\bibitem{FrHa}
Fredenhagen, K., Haag, R.: {\em Generally covariant quantum
field theory and scaling limits.} Commun.\ Math.\ Phys.\
{\bf 108} (1987) 91
%
\bibitem{Hes}
He{\ss}ling, H.: {\em  On the quantum equivalence principle.}
Nucl.\ Phys.\ {\bf B415} (1994) 243
%
\bibitem{Wol}
Wollenberg, M.: {\em Scaling limits and type of local algebras
over curved spacetime.} in:
{\em Operator algebras and topology.} Arveson, W.B. et al., eds.
                                      Pitman Research Notes
in Mathematics 270. Harlow: Longman 1992
%
\bibitem{ONei}
O'Neill, B.: {\em Semi-Riemannian geometry.} New York:
Academic Press 1983
%
\bibitem{Kay}
Kay, B.S.: {\em Linear spin-zero quantum fields in external
gravitational and scalar fields II.}
Commun. Math. Phys. {\bf 71} (1980) 29
%
\bibitem{BuDALo}
Buchholz, D., D'Antoni, C., Longo, R.: {\em Nuclear maps
and modular structures II.} Commun.\ Math.\ Phys.\ {\bf 129}
(1990) 631
%
\end{thebibliography}
\end{document}